\documentclass[prl,twocolumn,superscriptaddress]{revtex4-1}
\usepackage{graphicx}
\usepackage{gensymb}
\usepackage{amsmath}
\usepackage{amssymb}
\usepackage{mathrsfs}
\usepackage{xcolor}
\usepackage{multirow}
\usepackage[normalem]{ulem}
\usepackage{cancel}
\usepackage{siunitx}
\usepackage{graphicx}
\usepackage{dcolumn}
\usepackage{bm}

\begin{document}

\title{Myosin-independent amoeboid cell motility}

\author{Winfried Schmidt}
\affiliation{Theoretische Physik, Universit\"at Bayreuth, 95440 Bayreuth, Germany}
\affiliation{Univ. Grenoble Alpes, CNRS, LIPhy, F-38000 Grenoble, France
}
\author{Walter Zimmermann}
\affiliation{Theoretische Physik, Universit\"at Bayreuth, 95440 Bayreuth, Germany}
\author{Chaouqi Misbah}
\affiliation{Univ. Grenoble Alpes, CNRS, LIPhy, F-38000 Grenoble, France
}\author{Alexander Farutin}

\affiliation{Univ. Grenoble Alpes, CNRS, LIPhy, F-38000 Grenoble, France
}

\date{\today}

\begin{abstract}
Mammalian cell polarization and motility are important processes involved in many physiological and pathological phenomena, such as embryonic development, wound healing, and cancer metastasis. 
The traditional view of mammalian cell motility suggests that molecular motors, adhesion, and cell deformation are all necessary components for mammalian cell movement. However, experiments on immune system cells have shown that the inhibition of molecular motors does not significantly affect cell motility.
We present a new theory and simulations demonstrating that actin polymerization  
alone is sufficient to induce  spontaneously 
cell polarity accompanied by the retrograde flow. These findings provide a new understanding of the fundamental mechanisms {of cell movement and at the same time provide a simple mechanism for cell motility} in diverse configurations, e.g. on an adherent substrate, in a non-adherent matrix, or in liquids. 
\end{abstract}

\maketitle

{M}etazoan cell migration is often described as resulting from the polymerization of the actin network combined with its contraction by molecular motors, such as myosin. 
This type of cell migration is known as amoeboid mode and is characterized by a cell-scale flow of the cell cortex (actin filaments) \cite{Theriot1991}.
Actin filaments (known as F-actin) grow by polymerization of actin monomers (also known as G-actin, ''G'' for globular) at the cell front, inside the lamellipodium, and move as a crosslinked meshwork along the membrane towards the cell rear, where they depolymerize back to G-actin. G-actin monomers diffuse within the cytoplasm, reaching the cell front, where they polymerize again. 
Cell propulsion can be achieved by transmembrane proteins, such as integrins, which are anchored to the cortex and transmit the forces necessary for locomotion to the extracellular environment.

Myosin motors are believed to be key in force transduction, contracting the actin network and communicating internal forces to the substrate, resulting in forward propulsion of the cell. Retrograde flow of actin has been shown to drive amoeboid cell migration in various extracellular environments, including adhesion-free 3D matrix and tortuous geometry, 2D confinement, next to a non-adhesive substrate, or in bulk fluids \cite{lammermann2008rapid,Reversat2020,Shih_Yamada:2010.1,Ruprecht_2015_CCT,bagchi2002,AOUN20201157,barry2010dictyostelium,ONEILL2018}.

Theoretical models predict spontaneous cell polarization and motility for 2D, adhesion-based (known as mesenchymal) migration \cite{kozlov2007model, CallanJones2008, ziebert2011model, PhysRevLett.107.258103, BlanchMercader2013}.
Models for amoeboid migration explain cell polarization and motility through myosin motors \cite{hawkins2011spontaneous,recho2013contraction,Voituriez2016,Farutin2019,Mietke2019b}.
However, experiments on T-lymphocytes have shown that retrograde flow is induced even in the absence of myosin motor activity \cite{AOUN20201157}. Inhibiting myosin activity does not prevent polarization or perturb the retrograde flow significantly. By contrast, preventing actin polymerization significantly reduces retrograde flow velocity, and inhibiting both actin polymerization and myosin activity completely prevents cell polarization. This indicates that actin polymerization is the primary actor in amoeboid cell motility \cite{AOUN20201157}.  

Here we present  a model for cell polarization that does not require myosin activity, cell deformation, or adhesion to a substrate. First, we present a qualitative justification of the polarization mechanism on the molecular level and introduce the key ingredients of the model. Then we describe a more quantitative continuum-limit model of the cell-scale actin dynamics.
Using this model, we find analytically that there is a critical polymerization velocity above which the distribution of actin filaments at the cell membrane undergoes a spontaneous symmetry breaking, leading to cell polarization and the emergence of a self-sustained retrograde flow. The onset of cell polarization is either smooth (supercritical bifurcation) or hysteretic (subcritical bifurcation) in different parameter ranges. These analytical results for small amplitudes of the polarization and retrograde flow are confirmed by numerical simulations. Our simulations also show that cell polarization is accompanied by noticeable deformation of the originally spherical cell for strongly polarized cells.

\paragraph{ Basic mechanism of spontaneous cell polarization via actin polymerization.}

The actin filaments in the cell cortex undergo a continuous treadmilling process, wherein actin is polymerized at one end of the filament and depolymerized at the other end.
Due to the Brownian motion of filaments and membrane fluctuations, interstitial spaces constantly open up  between filaments and the membrane, allowing actin monomers to attach to filaments and polymerize, a process known as the Brownian ratchet mechanism \cite{MogilOster:96.1}. This polymerization near the membrane forces the filaments away from the membrane against viscous friction by the cytosol or other filaments, resulting in a net flow of actin filaments away from the membrane. The average orientation of this flow is perpendicular to the membrane.
In addition to this normal flow, the filaments can also slide along the membrane under the influence of other filaments or external factors. 

\begin{figure}
\begin{center}
\includegraphics[width=0.9\columnwidth]{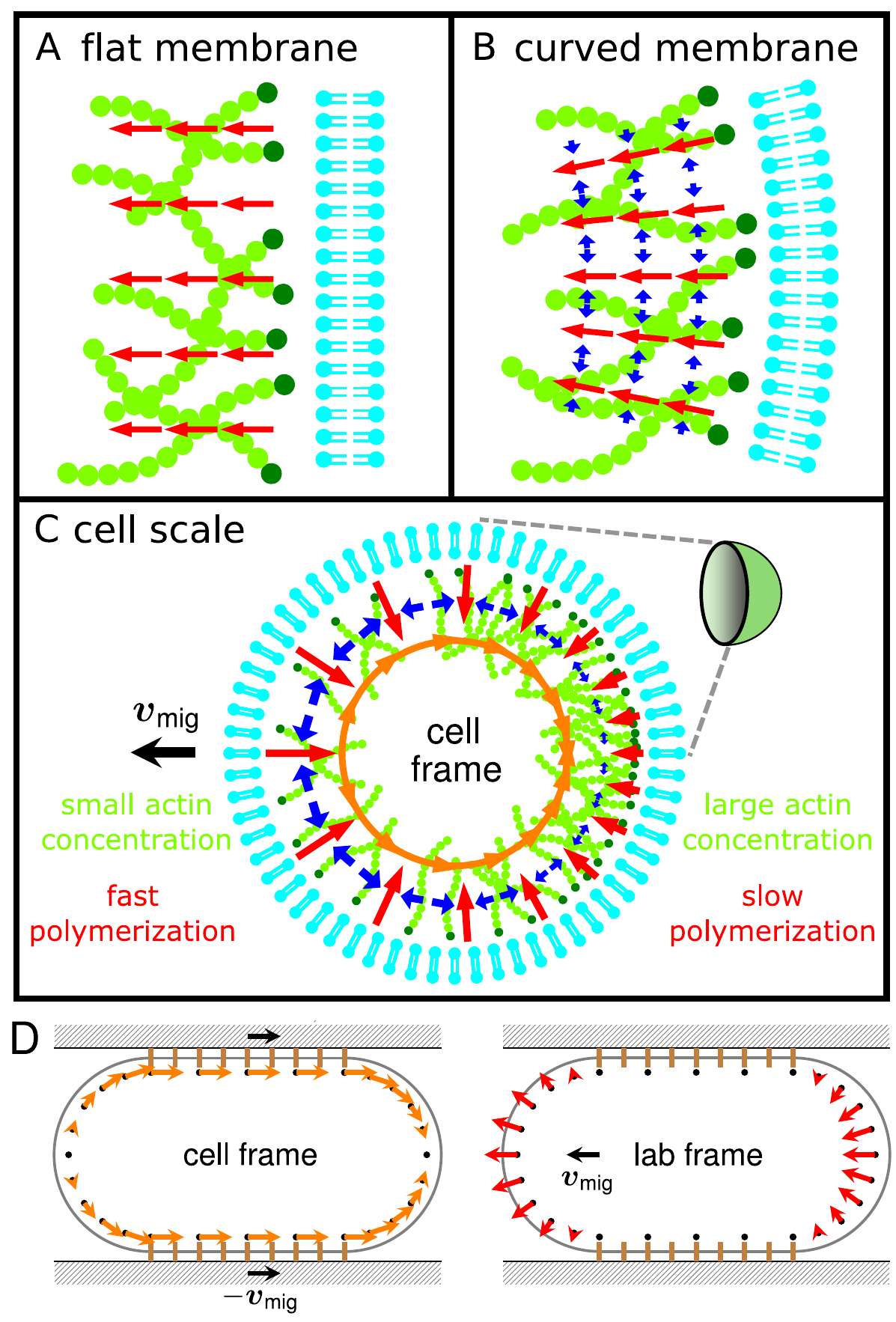}
\caption{
\label{fig1} Molecular mechanism of self-sustained cell polarity and retrograde flow: Cortical actin filaments (light-green chains) consisting of monomers (circles) polymerize (newly polymerized material shown in dark-green), leading to an inward treadmilling motion (red arrows indicate filament velocity) away from the membrane (turquoise).
A: No tangential stress is observed for a flat cortex.
B: Treadmilling normal to the membrane creates tangential viscous stresses (blue arrows) within a curved cortex.
C: Cross-section of the cell, showing the instability of a homogeneous filament distribution. The cortex stress (blue arrows) increases with increasing filament treadmilling velocity (red arrows), which in turn increases with decreasing filament concentration  due to a competition for monomers. Orange arrows show cell-scale flow of actin caused by inhomogeneous stresses, black arrow indicates migration direction.
D: Cross-section of an elongated cell migrating between two planar plates (dashed rectangles). Left (right) panel shows the polymerization velocity in the cell (laboratory) frame at reference points along the cell surface (black dots). Brown bars show transmembrane proteins.
}
\end{center}
\end{figure}

A key element of our model is that a curved membrane leads to a decrease of the distances between the depolymerizing ends of the filaments compared to the distances between their polymerizing ends. This effect is illustrated in Fig.\ref{fig1}A showing the flow of actin filaments away from a flat membrane that is stationary in the cell frame. Since the membrane is flat, this motion does not affect the relative distances among filaments within the cortex. However, for a convex curved membrane (Fig.\ref{fig1}B), the normal flow of the cell cortex due to actin polymerization brings actin units from different filaments closer to each other as they move away from the membrane. 
The cell cortex  consists of a cross-linked meshwork of actin filaments suspended in  cytosol, which generates viscous stresses. 
These stresses (blue arrows in Fig.\ref{fig1}B) are tangential to the membrane and push pairs of converging filaments apart.

Based on the above considerations, actin treadmilling, the process by which actin monomers are added to and removed from the filament ends, leads to the generation of tangential cortical stresses. These stresses are proportional to the membrane curvature and the actin polymerization rate. 
Motivated by experiments \cite{colin2023recycling}, we assume that the actin polymerization rate decreases with the concentration of actin filaments.
Therefore, the tangential stress is lower in the region of higher filament concentration than in the region of low actin concentration. This stress inhomogeneity acts as a pressure difference, pushing the cortex to slide along the membrane from the high-stress region to the low-stress region (motion direction shown by orange arrows in Figs.\ref{fig1}C and D (left)). This results in a tangential flow of the cortex (retrograde flow) in the cell frame, directed from the low-concentration region to the high-concentration one, which further reinforces the inhomogeneity of actin filament concentration and thus represents a positive feedback loop.
This translates to a normal flow in the laboratory frame which leads to membrane protrusion at the cell front and retraction at the rear, as shown in Fig.\ref{fig1}D (right).
This shows that the increase of the polymerization rate with decreasing actin filament concentration leads to an instability of the homogeneous distribution of cortical actin filaments.
We now propose  a thin shell continuum model describing this mechanism.

\paragraph{Thin shell model of the cell cortex.}
We model the actin cortex as a thin, 2D shell of {  viscous compressible fluid} that conforms to the membrane shape {  of the 3D cell}. 
Our model belongs to the general class of active gel models \cite{Kruse+.2004.1}, which have been successfully used to describe myosin-driven contraction of the cortex and the resulting motility \cite{Voituriez2016,Farutin2019}. However,  we deliberately exclude the activity of myosin motors to demonstrate that actin polymerization alone is sufficient to cause cell polarization. 
{  The viscous limit is justified since cell migration takes place at much larger times than the characteristic timescale of stress relaxation of the cortex  \cite{saha2016determining}.}

\begin{figure}
\includegraphics[width=0.98\columnwidth]{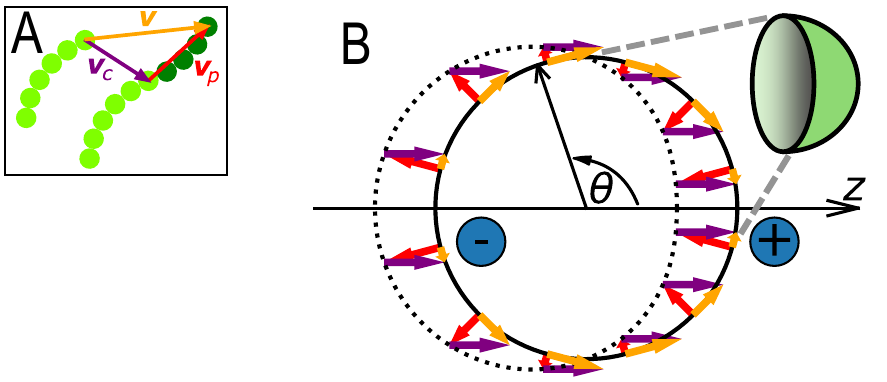}
\caption{\label{fig2} 
A: The filament displacement per unit time is described by the cortex velocity $\bm v_c$ (violet arrow), filament growth by the polymerization velocity $\bm v_p$ (red arrow), and motion of filament end points by the full velocity $\bm v$ (orange arrow).
B: Cross-section of the cell {(black solid circle)} showing polarity and retrograde tangential flow with {perturbation of the polymerization velocity (red arrows) and} full velocity $\delta {\boldsymbol v}$ ({orange} arrows).
Plus (minus) sign refers to the hemisphere with excess (reduced) concentration $c$.
{Violet arrows indicate cortex velocity perturbation $\delta \boldsymbol v_c = - \delta \boldsymbol v_\text{mig}$,}
{black dashed circle shows cell displacement after unit time.}
}
\end{figure}

In the continuum limit, the flow and treadmilling dynamics of the cortex are characterized by two velocity fields defined along the surface of the cell: the cortex velocity $\boldsymbol{v}_c$ and the full velocity $\boldsymbol{v}$. $\boldsymbol{v}_c$ refers to the locally averaged velocity of individual actin units that make up the filaments within the cortex. 
 It describes the local deformation of the cortex and is used to express the viscous response of the cortex. 
  Following existing thin shell models \cite{bacher2021three, Mietke2019, Farutin2019}, we obtain the cortex velocity by solving the cortex force balance (see below).
The full velocity refers to the locally averaged rate of change in position of the polymerizing ends of actin filaments adjacent to the membrane. Compared to $\boldsymbol{v}_c$, this velocity also includes the growth of the cortex via actin polymerization at the membrane, which is described by the locally averaged polymerization velocity $\boldsymbol{v}_p$: $\boldsymbol{v} = \boldsymbol{v}_c + \boldsymbol{v}_p$, see Fig.\ \ref{fig2}A. 
We denote the actin concentration, i.e., the number of polymerizing filaments per unit surface area, as $c$.
We assume that the polymerization velocity depends on $c$ as
\begin{equation}
\label{vp}
\boldsymbol{v}_p = v_p^0e^{-c/c_r}\boldsymbol{n},
\end{equation}
Herein, $v_p^0$ is the maximal polymerization velocity, $c_r$ a reference concentration, and $\boldsymbol{n}$ the unit vector normal to the membrane surface.

The passive viscous surface stress within the cortex, $\mathsf \sigma^s$, is determined by the gradients of the cortex velocity $\boldsymbol{v}_c$, $\mathsf \sigma^s=\eta_b(\boldsymbol\nabla^s\boldsymbol\cdot\boldsymbol v_c)\mathsf I^s+\eta_s[\boldsymbol\nabla^s\otimes\boldsymbol v_c\boldsymbol\cdot\mathsf I^s+\mathsf I^s\boldsymbol\cdot(\boldsymbol\nabla^s\otimes \boldsymbol v_c)^T]$.
Herein,  $\mathsf I^s = \mathsf I - \boldsymbol n \boldsymbol n$ is the operator of projection on the plane tangential to the membrane with unit matrix $\mathsf I$ and surface gradient operator $\boldsymbol\nabla^s\equiv \mathsf I^s\boldsymbol\cdot\boldsymbol\nabla$.
$\eta_s$ and $\eta_b$ are the shear and bulk 2D viscosities of the cortex.
$\mathsf \sigma^s$ satisfies the force balance equation $\boldsymbol\nabla^s\boldsymbol\cdot \mathsf \sigma^s+\boldsymbol f_t+\Delta P \boldsymbol n= \boldsymbol 0$, with
the tension force induced by the cell membrane, 
$\boldsymbol f_t=-H\zeta_0\boldsymbol n$ \cite{Farutin2019, Voituriez2016}, where $\zeta_0$ is the membrane tension and $H$ the mean curvature of the membrane.
$\Delta P$ is the hydrostatic pressure difference between the  inside and outside of the cell.
$\zeta_0$ is calculated by fixing the surface area of the membrane $A_0$.
The force balance allows us to compute the cortex velocity $\boldsymbol{v}_c$.
This determines $\boldsymbol{v}_c$ up to a constant velocity vector since the system is Galilei invariant (rigid-body translations and rotations do not deform the cortex).
The normal component of $\boldsymbol{v} = \boldsymbol{v}_c + \boldsymbol{v}_p$ determines the cell shape evolution. 
The filament concentration obeys \cite{SM}
 \begin{equation}
		\label{cdot_advective}
		\dot c+\boldsymbol\nabla^s\boldsymbol\cdot(\boldsymbol vc)=D\Delta^s c+\beta(c_0-c),
\end{equation}
with actin diffusion coefficient $D$, Laplace-Beltrami operator $\Delta^s$, restoration rate $\beta$, and homeostatic actin concentration $c_0$.

%

\paragraph{Cell polarization and the emergence of retrograde flow as dynamical instability.}
An unpolarized spherical cell {  with radius $R=R_0$} is a basic steady state {  of our model}, regardless of the parameters. {\color {black} It is} characterized by a homogeneous filament concentration $c=c_0$ along the cortex and no tangential flow of actin filaments.
A position-independent polymerization velocity $\boldsymbol v_p=v_p(c_0)\boldsymbol n$ 
is found to be fully canceled by an inward radial cortex velocity $\boldsymbol v_c=-v_p(c_0)\boldsymbol n$, where $v_p(c_0)=v_p^0e^{-c_0/c_r}$.
The vanishing sum $\boldsymbol v_p+\boldsymbol v_c\equiv\boldsymbol v= \bm 0$ implies  an absence of membrane deformation and of tangential cortex flow.

By increasing $v_p^0$ the homogeneous concentration $c_0$  becomes unstable with respect to an axisymmetric 
concentration fluctuation $\delta c(\boldsymbol r,t)= \delta \hat c_1 (t) \cos(\theta)$, where $\theta$ is the angle with respect to the polarization direction (Fig.\,{  \ref{fig2}B}). 
Substituting  this into  (\ref{vp}), gives $\boldsymbol v_p(\boldsymbol r,t)\approx v_p^0(c_0)(1-\delta c(\boldsymbol r,t)/c_r)\boldsymbol n(\boldsymbol r)$, yielding a polymerization velocity perturbation
\begin{equation}
\label{deltavp}
\delta \boldsymbol v_p=- \delta\hat v\cos(\theta)\, \boldsymbol n\,,
\quad \delta\hat v=\frac{v_p^0 \exp\left(-\frac{c_0}{c_r}\right) \delta \hat c_{1}}{c_r} \, ,
\end{equation}
which is normal to the cell shape.
$\delta {\boldsymbol v}_p$ points inwards (outwards) in the excess (reduced) concentration region, see red arrows in Fig.\ \ref{fig2}B.
Suppose that all membrane points move with $\delta {\boldsymbol v}_p$ relative to an extracellular environment, then this implies a migration of the cell center with velocity $\delta \boldsymbol{v}_\text{mig} = - \delta \hat v \boldsymbol e ^z$.
A Galilei transformation of $\delta {\boldsymbol v}_p$ from the laboratory to the cell frame then yields $\delta \boldsymbol v=\delta \boldsymbol v_p + \delta \boldsymbol v_c = - \delta \hat v \sin \theta \boldsymbol t^\theta$ (orange arrows in Fig.\ \ref{fig2}B) where  $\boldsymbol t^\theta$ is the unit vector in direction of increasing $\theta$. Herein, $\delta \boldsymbol v_c=\delta \hat v \boldsymbol e ^z$ is the constant cortex velocity perturbation (violet arrows in Fig.\,{\ref{fig2}B}).
The full velocity perturbation $\delta {\boldsymbol v} $ is purely tangential and therefore conserves the cell shape. This retrograde flow
advects actin from the low concentration pole (cell front) to the high concentration pole (cell rear). Actin diffusion  and {restoration of the concentration} act against this tendency, so that 
perturbations $\delta c$ ({cortical polarity})  and  $\delta \boldsymbol v$ self-amplify above the critical polymerization speed \cite{SM}
\begin{equation}
\label{vpc}
v_p^c=\frac{c_r}{c_0}\left(\frac{D}{R_0}+\frac{\beta R_0}{2}\right)e^{c_0/c_r}\,
\end{equation}
and grow exponentially as long as $\delta c$ is still small. 
For simulation results with $\beta = 0$ we refer to the SI \cite{SM}.
Eq.\ \eqref{vpc} also shows that  $v_p^c$ grows to infinity in the flat-membrane limit (very large $R_0$). 
{Higher-order harmonics  become  unstable for higher speeds than that of Eq.\ \eqref{vpc}, see \cite{SM}. }
In the supercritical region, $v_p>v_p^c$, nonlinearities
lead to saturation of $\delta c$. 

\begin{figure}
\includegraphics[width=\columnwidth]{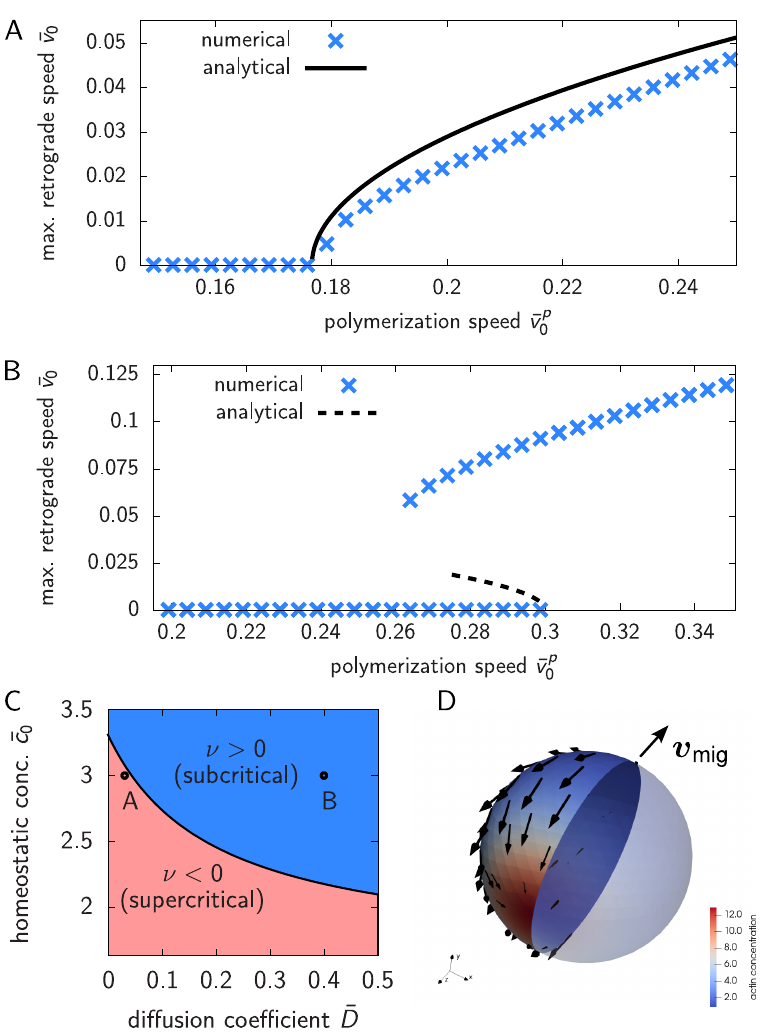}
\caption{\label{fig3}
A: The analytical (solid line) and numerical ($\times$) branches of the supercritical bifurcation of the retrograde flow velocity for $D=0.03$.
B: Both are compared for a subcritical bifurcation with $D=0.4$.  
C: The solid line corresponds to $\nu=0$ (boundary between super- and subcritical bifurcation).
D: Steady-state numerical solution of the cell
for $D = 0.4$ and $v_p^0 = 6$.
One hemisphere is shown transparent for illustration.
Color code (see legend) indicates cortical actin filament concentration, showing polarity.  
Black arrows along the surface show tangential retrograde flow which points opposite to migration direction $\boldsymbol{v}_\text{mig}$.
For further parameters see \cite{SM}.
}
\end{figure}

\paragraph{Finite cell polarization and retrograde flow velocity.} 
Close above the instability threshold, the cell polarization and  retrograde flow can be determined analytically by perturbation expansion \cite{Strogatz:94}, yielding
\begin{equation}
\label{WNLA}
\partial_t \delta \hat c_{1}=(\lambda_1+\nu \delta \hat c_{1}^2 )\delta \hat c_{1}.\
\end{equation}
$\lambda_1$ and 
$\nu$ are given in terms of model parameters in SI \cite{SM} (Eqs.\ (21) and (52) therein).
For $\nu<0$ the bifurcation  is supercritical as shown for the retrograde flow for $D=0.03$ in Fig.\,\ref{fig3}A.
In this case Eq.\,(\ref{WNLA}) has the steady-state solution $\delta \hat c_1^2=-\lambda_1/\nu$.
For  $\nu>0$, the bifurcation  is subcritical. The transition is hysteretic with two stable branches  (crosses in Fig.\,\ref{fig3}B).
The phase diagram in Fig.\,\ref{fig3}C shows  super- and subcritical bifurcations regions.

The numerical solution  is detailed in the SI \cite{SM}, showing reasonable agreement with analytical results (Fig.\,\ref{fig3}).
 Our numerical results confirm the proposed motility mechanism even for highly deformed cells well above $v_p^c$.

\paragraph{Conclusion}
The instability is independent of how the retrograde flow is transmitted to the external environment for cell migration, such as crawling on a substrate, swimming in a fluid, or adhesion-free migration in a  3D matrix. The migration velocity is of order of retrograde flow velocity \cite{AOUN20201157}, but its determination requires a precise analysis of motility.

Relation\ \eqref{vp} is supported by \textit{in vitro} experiments showing that filament growth is slowed due to competition for available monomers \cite{colin2023recycling}.
If $\mu$ is the number of diffusing G-actin monomers per unit volume and $h$ the cortex thickness, we define the dimensionless parameter $\bar{\mu} = \frac{\mu h}{c}$.
Assuming that $\bar{\mu}$ is small, the polymerization velocity reads  $v_p (\bar{\mu}) \approx V_1 \bar{\mu}$. 
In the SI \cite{SM} we compute $V_1$ based on Ref.\ \cite{colin2023recycling}.
Therein, slowed polymerization takes place for a ratio of $\bar{\mu} = 0.61$.
For migrating fibroblasts, $\mu = \SI{8.5}{\micro \mol \per \liter}$ and $c = \SI{1370}{\per \micro \meter \squared}$ are reported \cite{ABRAHAM19991721}, yielding a comparable value of $\bar{\mu} = 0.71$.
This indicates that filament polymerization can be potentially limited by a competition for monomers in living cells.
This  is further supported by the observation that a large percentage of diffusing actin in living cells is in the form of short oligomers and non-polymerizable monomeric G-actin \cite{aroush2017actin}.
Our estimate suggests a relation $v_p \propto  c^{-1}$. 
	In Eq.\ \eqref{vp} we assume an exponential relation,  ensuring numerical stability even for small  $c$.
	Still, we find similar behavior for the migration speed for both  laws \cite{SM}.

For simplicity, we neglected here the elastic resistance of the cortex \cite{Farutin2019}. The associated passive stresses promote the homogenization of the actin concentration and thus counter-act the  instability. In the SI \cite{SM} we derive the corresponding instability threshold which lies in the range of typical polymerization speeds.

Our thin shell model can be rewritten using only a single velocity field $\boldsymbol{v}$ \cite{SM}. This yields an active surface stress $\sigma^s_\text{active}$ caused by filament polymerization which is proportional to $v_p^0$ and the membrane curvature (Fig.\ \ref{fig1}B and C).
Computing $\sigma^s_\text{active}$ shows that our thin shell equations are consistent with a full 3D description of the cortex, yielding $\eta_s = h \mu$ and $ \eta_b = \frac{2 h \lambda \mu}{\lambda + 2 \mu}$ with the shear and bulk 3D cortical viscosities $\mu$ and $\lambda$ \cite{SM}.

It has been suggested that the interplay between the polymerization  and the cell shape could lead to shape polarization \cite{BlanchMercader2013}. 
However, the polymerization velocity was assumed to be constant and motility arises only in the nonlinear regime.
Another model considered actin treadmilling as a passive background process, leading to a fingering instability of the cell shape due to myosin contractility \cite{CallanJones2008}.
However, the first harmonic, responsible for actin polarization, is neutral, precluding the emergence of motility. 
Another study coupled the cortex forces driving the growth of cell membrane protrusions to the concentration of a chemical species distributed within the cytosol \cite{Lavi2020}. This model shows that spontaneous polarization, motility, and deformation can occur for a cell confined between two plates. 
However, actin which freely diffuses in the cytosol is in monomer form.
In our case, the active species resides only in the cortex, coinciding with the filaments themselves.
Another model \cite{John2008} describes actin-polarization-driven motility   for   the listeria bacterium \cite{Theriot1992,weiner,sykes1,sykes2,sykes3,GERBAL20002259}, but the actin network is taken to be elastic, preventing retrograde flow.

Nematode sperm cells rely on the major sperm protein (MSP) for motility without molecular motors \cite{roberts1997nematode, bottinoNematode}. 
Similar to actin, MSP polymerizes at the cell front 
and depolymerizes at the rear, creating
a cell-scale flow of the MSP gel. 
The proposed mechanism relies on a prescribed pH gradient across the lamellipodium \cite{bottinoNematode} and is thus inherently different from the spontaneous symmetry breaking described by our model.

We would like to thank J. \'Etienne, {  K. John,} P. Recho, and O. Th\'eodoly for many stimulating discussions of cellular motility and R. Voituriez for early feedback on this work. 
W. S. thanks for the financial support of the DAAD and the study program Biological Physics of the Elite Network of Bavaria.
W. S., A. F., and C. M. thank CNES (Centre National d’Etudes Spatiales) for a financial support.
All authors thank the French-German university program ”Living Fluids” (grant CFDA-Q1-14).

\end{document}


\title{
	Myosin-independent amoeboid cell motility \\
	-- Supplementary information --
}

\begin{abstract}
This document contains {further explanation on our model equations}, a derivation of our thin shell model using the active stress,
the general form of the linear stability analysis of the homogeneous unpolarized solution, the calculation of the retrograde flow velocity in the weakly nonlinear regime, the numerical method for solving the full problem, the parameter values used in numerical and analytical calculations, an estimate for the competition of polymerizing actin filaments for monomers, additional results for the case of zero restoration of the actin concentration,
{a linear stability analysis considering a finite elastic modulus of the cortex, and a weakly nonlinear analysis using an inverse polymerization law.}
\end{abstract}

\author{Winfried Schmidt}
\affiliation{Theoretische Physik, Universit\"at Bayreuth, 95440 Bayreuth, Germany}
\affiliation{Univ. Grenoble Alpes, CNRS, LIPhy, F-38000 Grenoble, France}
\author{Walter Zimmermann}
\affiliation{Theoretische Physik, Universit\"at Bayreuth, 95440 Bayreuth, Germany}
\author{Chaouqi Misbah}
\affiliation{Univ. Grenoble Alpes, CNRS, LIPhy, F-38000 Grenoble, France}
\author{Alexander Farutin}
\affiliation{Univ. Grenoble Alpes, CNRS, LIPhy, F-38000 Grenoble, France}

\date{\today}
\maketitle
\onecolumngrid
\setlength{\parindent}{0cm}

\section{Form of the advection diffusion equation} 
The full velocity field $\boldsymbol v$  determines the shape evolution of the membrane and therefore of the cell, and 
it determines also the advection of the filament concentration $c$ via the following equation:
\begin{equation}
\label{cdot}
\dot c+(\boldsymbol\nabla^s\boldsymbol\cdot \boldsymbol v)c=\beta(c_0-c)+D\Delta^s c.
\end{equation}
The term $(\boldsymbol\nabla^s\boldsymbol\cdot \boldsymbol v)c$ corresponds to the change of concentration due to the local dilatation of the cortex, where $(\boldsymbol\nabla^s\boldsymbol\cdot\boldsymbol v)$ is equal to the dilatation rate.
The term $\beta(c_0-c)$ restores the concentration to the homeostatic value $c_0$, where $\beta$ is the rate at which a given filament can stop treadmilling, due to, for example, attachment of a capping protein.
The term $D\Delta^s c$ refers to the diffusion of actin filaments along the cell surface, where $D$ is the diffusion coefficient and $\Delta^s c=\boldsymbol\nabla^s\boldsymbol\cdot(\boldsymbol\nabla^s c)$.
$\dot c$ in Eq. (\ref{cdot}) refers to {the material derivative of the concentration, i.e.,} the evolution of the concentration at the position of a given actin filament {(Lagrangian description)}.
This equation is valid regardless of the normal deformation or tangential flow of the cortex.
If the shape of the cell is constant, it may be more convenient to track the evolution of the concentration at a given point of the cell shape {(Eulerian description)}.
In this case, the left hand side of Eq. (\ref{cdot}) transforms into a more traditional advective form, 
\begin{equation}
	\label{cdot_advective}
	\dot c+\boldsymbol\nabla^s\boldsymbol\cdot(\boldsymbol vc)=\beta(c_0-c)+D\Delta^s c,
\end{equation}
{which corresponds to the form of the advection-diffusion equation used in the main text [Eq.\ (2) therein].}
{ 
	\section{Active stress and derivation of the thin shell model} \label{secFiniteThickness} }
The approach chosen for the analytical calculations is to decompose the full velocity field at the membrane of the cell into the polymerization part and the cortex deformation part {which yields the passive viscous surface stress
\begin{equation}
	\label{passive}
	\sigma^s=\eta_b(\boldsymbol\nabla^s\boldsymbol\cdot\boldsymbol v_c)\mathsf I^s+\eta_s[\boldsymbol\nabla^s\otimes\boldsymbol v_c\boldsymbol\cdot\mathsf I^s+\mathsf I^s\boldsymbol\cdot(\boldsymbol\nabla^s\otimes \boldsymbol v_c)^T].
\end{equation}
}
We could also take an alternative approach, writing Eq. (\ref{passive}) in terms of the full velocity field $\boldsymbol v$.
Substituting $\boldsymbol v_c=\boldsymbol v-\boldsymbol v_p$ into Eq. (\ref{passive}) allows us to rewrite the whole model in terms of a unique velocity field $\boldsymbol v$.
This comes at a price of an additional "active" contribution to the stress tensor that is related to the gradients of the polymerization velocity:
Writing the surface gradient for a normal velocity field {  $\bm v_p (\bm r) = v_p (\bm r) \bm n (\bm r)$} yields
{ 
\begin{equation}
	\label{activestress1}
	\bm \nabla^s \bm \cdot (v_p \bm n) = \bm \nabla^s v_p \bm \cdot \bm n + v_p \bm \nabla^s \bm \cdot \bm n = v_p H
\end{equation}
and}
\begin{equation}
\label{activestress2}
\boldsymbol\nabla^s\otimes(v_p\boldsymbol n)\boldsymbol\cdot\mathsf I^s =v_p\boldsymbol\nabla^s\otimes\boldsymbol n\boldsymbol\cdot\mathsf I^s+\boldsymbol\nabla^sv_p\otimes\boldsymbol n\boldsymbol\cdot\mathsf I^s=v_p\boldsymbol\nabla^s\otimes\boldsymbol n.
\end{equation}
Here $\boldsymbol\nabla^s\otimes\boldsymbol n$ is a symmetric tensor, with components only along the tangent directions to the cell surface.
Its two non-zero eigenvalues correspond to the principal curvatures of the membrane.
{ Using Equations \eqref{activestress1} and \eqref{activestress2}, one obtains from Equation \eqref{passive}
\begin{equation}
	\label{activestress}
	\sigma^s=\eta_b(\boldsymbol\nabla^s\boldsymbol\cdot\boldsymbol v)\mathsf I^s+\eta_s[\boldsymbol\nabla^s\otimes\boldsymbol v\boldsymbol\cdot\mathsf I^s+\mathsf I^s\boldsymbol\cdot(\boldsymbol\nabla^s\otimes \boldsymbol v)^T] + \sigma^s_\text{active}
\end{equation}
with the active viscous surface stress
\begin{equation}
	\label{activestress3}
	\sigma^s_\text{active} = - v_p \left(\eta_b  H \mathsf I^s + 2 \eta_s \boldsymbol\nabla^s\otimes\boldsymbol n \right).
\end{equation}
}
Equation \eqref{activestress3} thus shows that the effective active stress due to the treadmilling of actin filaments is tangential and proportional to the treadmilling velocity and the curvature of the cell.
This equivalent reformulation relates the thin-shell model to the qualitative model presented in Fig. 1 of the main text.

{ 
In the following, we show that our thin shell equations are consistent with a full 3D description of a cortex with finite thickness.
To this end, we solve the compressible Stokes equation in a 3D domain for an exemplary cylindrical geometry.
From this we obtain the 3D active viscous stress which, upon integration over the thickness of the cortex, yields the surface stress.
Taking the limit of small cortical thickness, this yields explicit expressions for the 2D shear and bulk viscosities used in our thin shell model.} 

{ We consider a fixed, cylindrical cell shape with radius $R_0$ and infinite extension along the axial direction, see Fig.\ \ref{fig_cyliner}.
\begin{figure}[h]\centering
	\includegraphics[width=0.33\textwidth]{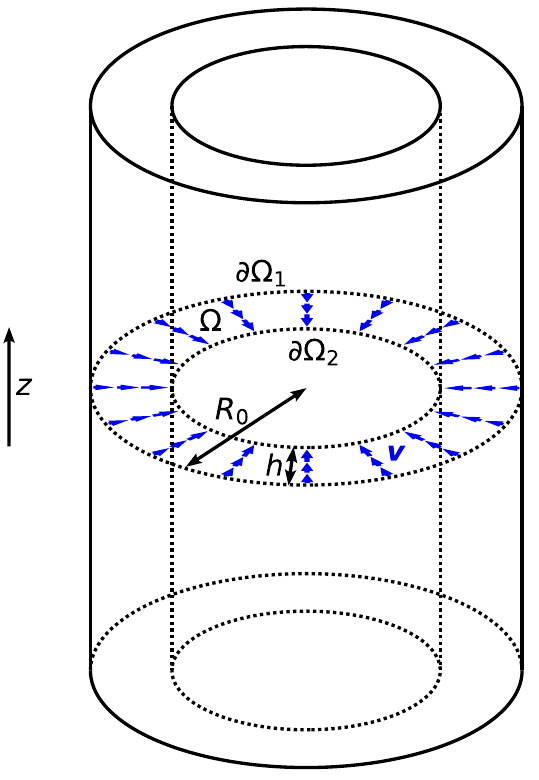}
	\caption{{  3D model for the cortex with finite thickness $h$ in a cylindrical cell geometry with radius $R_0$ and infinite extension along the $z$-direction. The cortex domain is denoted as $\Omega$ where $\partial \Omega_1$ and $\partial \Omega_2$ refer to its outer and inner boundary, respectively. Blue arrows indicate the cortex velocity $\bm v$.}}
	\label{fig_cyliner}
\end{figure}
Using cylindrical coordinates $(r , \varphi , z )$, the cortex domain is denoted as $\Omega = \{ \bm r \in \mathbb{R}^3 | R_0 - h < r < R_0 \}$ where $h$ is the constant thickness of the cortex.
In the following, the outer edge of the cortex ($r = R_0$) is denoted as $\partial \Omega_1$ and the inner edge ($r = R_0 - h$) as $\partial \Omega_2$. 
We neglect interactions of the cortex with the surrounding plasma membrane and with the enclosed cytoplasm.
The problem is formulated as 

\begin{align}
	\label{eqsThinShellCylinderForceBalance}
	\bm \nabla  \cdot \sigma &= \bm 0 \quad \text{in} \ \Omega, \\
	\label{eqsThinShellCylinderBC1}
	\bm v &= - v_p \bm e_r \quad \text{on}  \ \partial \Omega_1, \\
	\label{eqsThinShellCylinderBC2}
	\sigma  \cdot \bm e_r &= \bm 0 \quad \text{on} \ \partial \Omega_2.
\end{align}
Equation \eqref{eqsThinShellCylinderForceBalance} is the force balance with $\bm \nabla$ being the 3D Nabla operator and 
\begin{equation}\label{eq3DStress}
	\sigma = \lambda \bm \nabla \cdot \bm v \mathsf I + \mu \left[ \bm \nabla \otimes \bm v + \left( \bm \nabla \otimes \bm v \right)^T \right]
\end{equation}
is the 3D active viscous stress.
Herein, $\lambda$ and $\mu$ denote the 3D bulk and shear viscosity, respectively, and $\mathsf I$ is the identity matrix. 
The first boundary condition \eqref{eqsThinShellCylinderBC1} fixes the velocity at the outer edge of the cortex where $v_p$ is a constant polymerization speed and $\bm e_r$ is the unit vector in radial direction.
As a second boundary condition we assume in equation \eqref{eqsThinShellCylinderBC2} a force-free inner edge of the cortex.

The symmetry of the problem dictates that $\bm v (r, \varphi) = v_r (r) \bm e_r (\varphi)$.
From Equation \eqref{eqsThinShellCylinderForceBalance} then follows
\begin{equation}\label{eqCauchyEuler}
	(\lambda + 2 \mu) \partial_r^2 v_r + \frac{\lambda + 2 \mu }{r} \partial_r v_r - \frac{\lambda + 2 \mu }{r^2} v_r = 0.
\end{equation}
Together with Eqs.\ \eqref{eqsThinShellCylinderBC1} and \eqref{eqsThinShellCylinderBC2} one obtains
\begin{equation}\label{eqvR}
	v_r (r) = - \frac{v_p R_0 \left[ (\lambda + \mu) (R_0 - h)^2 + \mu r^2 \right]}{r \left[ (\lambda + \mu) (R_0 - h)^2 + \mu R_0^2 \right]}.
\end{equation}
Substitution into Eq.\ \eqref{eq3DStress} yields expressions for $\sigma_{rr} (r)$, $\sigma_{\varphi \varphi} (r)$, and $\sigma_{zz} (r)$ (note that off-diagonal elements of the stress tensor are zero).
The surface stresses $\sigma_{ij}'$ are then obtained by integration over the thickness of the cortex,
\begin{equation}\label{eqStressesIntegrated}
	\sigma_{ij}' = \int_{R_0 - h}^{R_0} \sigma_{ij} (r) \ dr.
\end{equation}
Assuming that the thickness of the cortex is much smaller than the cell radius, i.e., $h \ll R_0$, one obtains from Eq.\ \eqref{eqStressesIntegrated} up to linear order 
\begin{equation}\label{eqStressesLinear}
	\sigma_{rr}' = \mathcal{O} (h^2), \qquad \sigma_{\varphi \varphi}' = - \frac{4 v_p h \mu (\lambda + \mu)}{R_0 (\lambda + 2 \mu)} + \mathcal{O} (h^2), \qquad \sigma_{zz}' = - \frac{2 v_p h \mu \lambda}{R_0 (\lambda + 2 \mu)} + \mathcal{O} (h^2).
\end{equation}
These active surface stresses which are obtained from the full 3D calculation can now be compared to the surface stresses of our thin shell model. Eq.\ \eqref{activestress3} yields for the cylindrical geometry
\begin{equation}\label{eqSurfaceStressesThinShell}
	\sigma^s_{\text{active}, rr} = 0, \qquad \sigma^s_{\text{active}, \varphi \varphi} = - \frac{v_p (\eta_b + 2 \eta_s)}{R_0}, \qquad \sigma^s_{\text{active}, zz} = - \frac{v_p \eta_b}{R_0},
\end{equation}
and $\sigma^s_{\text{active}, ij} = 0$ for $i \neq j$.
Comparison of Eqs.\ \eqref{eqStressesLinear} and \eqref{eqSurfaceStressesThinShell} yields
\begin{equation}\label{eq2DViscosities3D}
	\eta_b = \frac{2 h \lambda \mu}{\lambda + 2 \mu}, \qquad \eta_s = h \mu.
\end{equation}
Eq.\ \eqref{eq2DViscosities3D} shows that the surface bulk and shear viscosities are functions of the 3D viscosities. This provides a direct link between the thin shell equation \eqref{activestress} and the 3D active viscous stress in Eq.\ \eqref{eq3DStress}.
We also performed the equivalent calculation for a spherical cell shape. From this we obtain the relation
\begin{equation}\label{eqViscositiesSphere}
	\eta_b + \eta_s = \frac{h \mu (3 \lambda + 2 \mu)}{\lambda + 2 \mu}
\end{equation}
which is consistent with Eq.\ \eqref{eq2DViscosities3D}. }

\section{Linear stability analysis for actin polarization on a spherical shape.}
\label{sec_linear_stability_1}
The basic solution is a homogeneous spherical cell and we show here that actin concentration gets spontaneously polarized beyond a critical polymerization speed.
Unpolarized spherical cell is a trivial steady-state solution of the above model, regardless of the parameters.
This solution is characterized by a homogeneous concentration of actin filaments $c=c_0$ along the inner surface of the membrane and no tangential flow of actin filaments.
Position-independent polymerization velocity $\boldsymbol v_p=v_p(c_0)\boldsymbol n$ is fully canceled by an inward radial flow of cortex $\boldsymbol v_c=-v_p(c_0)\boldsymbol n$, where $v_p(c_0)=v_p^0e^{-c_0/c_r}$.
The vanishing sum $\boldsymbol v_p+\boldsymbol v_c\equiv\boldsymbol v= \boldsymbol0$ implies absence of membrane deformation and of tangential cortex flow.
Substituting the cortical velocity $\boldsymbol v_c=-v_p(c_0)\boldsymbol n$ into Eq. (\ref{passive}) and {the force balance equation (see main text)} 
we see that actin treadmilling generates an additional pressure that pushes against the membrane.
This pressure is independent of the position along the surface of cell and is added to the osmotic pressure $\Delta P$.
Together they are compensated by the membrane tension.

We find that this spherically symmetric basic state becomes unstable against perturbations of the filament concentration, $\delta c({\bf r},t)$, when the polymerization velocity is increased beyond a critical value $v_p^c$.
For the determination of $v_p^c$ we separate the inhomogeneous actin filament concentration $\delta c({\bf r},t)$ by using the ansatz $c({\bf r},t)=c_0+\delta c({\bf r},t)$.
At first, we assume small $\delta c$,  restrict for simplicity our analysis to axisymmetric $\delta c({\bf r},t)$ and choose $\delta c({\bf r},t)= \delta \hat c_1(t) P_1(\theta)$ with the first spherical harmonic $P_1(\theta)=\cos(\theta)$, where $\theta$ is the angle with respect to the polarization direction, which we choose as $z$-axis (see main text).
$\delta \hat c_1(t)$ is the time-dependent amplitude.
Note that small perturbations of the first harmonic of $\delta c({\bf r},t)$ do not change the spherical cell shape at the leading order, i.e., the instability to cell polarization occurs in spherical cells.
The linear stability analysis below in section \ref{sec_linear_stability_supplement} shows that basic state becomes first unstable with respect the first harmonic perturbation. More precisely, the instability occurs at lower $v_p^c$ than for the higher-order harmonics perturbations.

According to {the expression for $\boldsymbol{v}_p (c)$ (see main text)}, a small $\delta c$ causes a perturbation of the polymerization velocity via $\boldsymbol v_p(\boldsymbol r,t)\approx v_p^0(c_0)(1-\delta c(\boldsymbol r,t)/c_r)\boldsymbol n(\boldsymbol r)$ with the perturbation $\delta \boldsymbol v_p$ written in the following form:
\begin{equation}
\label{deltavp}
\delta \boldsymbol v_p=- \delta\hat v\cos(\theta)\, \boldsymbol n\,,
\quad \delta\hat v=\frac{v_p^0 \exp\left(-\frac{c_0}{c_r}\right) \delta \hat c_{1}}{c_r} \,.
\end{equation}
This relationship again illustrates that an increasing $c$ decreases $v_p$ and vice versa.
The position-dependent perturbation of the polymerization velocity in Eq.\,(\ref{deltavp}), indicated by the red arrows in Fig. 2 (main text), is directed inward at the pole $\theta=0$ of the cell and outward at the pole $\theta=\pi$.
For both poles, the perturbation $\delta {\boldsymbol v}_p $ points in the $-z$ direction.
Since the normal component of $\delta \boldsymbol v= \delta \boldsymbol v_p +\delta \boldsymbol v_c$ vanishes at the surface of the resting cell because of the shape conservation, the polymerization velocity is compensated at both poles by the perturbation of the cortex velocity, which, in general, must be calculated by solving the force balance equation {(see main text)}.
We found that this solution corresponds to a position-independent velocity $\delta \boldsymbol v_c= \delta\hat v  {\boldsymbol e}^z$ (blue arrows in Fig. 2 of the main text), which does not affect the force-balance equation {(see main text)} because it does not generate any deformation in the cortex.
One can see that the normal component of the constant $\delta {\boldsymbol v}_c$, $(\delta {\boldsymbol v}_c \cdot {\boldsymbol n}){\boldsymbol n}({\boldsymbol r})=\delta\hat v \cos(\theta) {\boldsymbol n} ({\boldsymbol r})$, compensates at each point along the sphere $\delta {\boldsymbol v}_p$ given by Eq.\,(\ref{deltavp}).
It is important to note that only the normal part of $\delta\boldsymbol v_c$ is canceled by $\delta \boldsymbol v_p$ when calculating $\delta\boldsymbol v$.
The velocity $\delta\boldsymbol v$ is thus equal to the tangential component of $\delta {\boldsymbol v}_c$, which describes the retrograde flow along the cell surface (green arrows in Fig. 2 of the main text),
\begin{equation}
\label{infinitan}
\delta \boldsymbol v=-\delta\hat v\sin (\theta) \boldsymbol{t}^\theta\, ,
\end{equation}
where $\boldsymbol{t}^\theta$ is the unit tangent vector along the polar direction defined as $\boldsymbol{t}^\theta=(\cos(\theta) \cos(\varphi), \cos(\theta)\sin(\varphi), -\sin(\theta))$ and $\varphi$ is the azimuthal angle.
Since the angular dependence of the $z$-component  of the retrograde flow (\ref{infinitan}) is $\delta { v}_z=\delta\hat v \sin^2\theta>0$, the actin flows along the cell surface from the $-z$ (low concentration pole) to the $+z$ (high concentration pole) region.
It is a key finding that an actin concentration fluctuation with amplitude $\delta \hat c_1\cos\theta$ generates a polymerization-induced retrograde flow in a spherical cell.
More importantly, this retrograde flow amplifies the initial perturbation $\delta c$ by bringing actin filaments from the low-concentration pole to the high-concentration one.

We now substitute the perturbation $\delta {\boldsymbol v}= \delta {\boldsymbol v}_p+\delta {\boldsymbol v}_c$ from Eq. (\ref{infinitan}) into Eq.\,(\ref{cdot_advective}) and extract the leading term (coefficient of $\cos\theta$).
This gives the evolution equation for $\delta\hat c_1$ as $\partial_t\delta \hat c_{1}=\lambda_1\delta \hat c_{1}$ with $\lambda_1$ given by
\begin{equation}
\label{lambda1}
\lambda_1 = 2\frac{v^0_p c_0}{R_0 c_r} e^{-\frac{c_0}{c_r}}- 2\frac{D}{R_0^2}  - \beta.
\end{equation}
The linear stability coefficient $\lambda_1$ results from a competition between the advection term that tends to amplify the perturbation and the terms in the right hand side of Eq. (\ref{cdot_advective}) that tend to homogenize the distribution of actin filaments in the cortex.
The onset of cell polarization takes place at $\lambda_1=0$.
This condition provides the critical polymerization velocity:
\begin{equation}
\label{vpc}
v_p^c=\frac{c_r}{c_0}\left(\frac{D}{R_0}+\frac{\beta R_0}{2}\right)e^{c_0/c_r}\,.
\end{equation}
This analytical expression for $v_p^c$ explicitly shows the parameter dependence for the threshold to the onset of cell polarization and the retrograde flow and, in particular, also its crucial dependence on the cell curvature ($v_p^c$ grows to infinity in the flat-membrane limit, given by infinitely large $R_0$). 
In the supercritical region, $v_p>v_p^c$, one finds self-amplification of small perturbations $\delta c$ with respect to the homogeneous concentration $c_0$.
Then $\delta c$ and the associated velocity fields grow exponentially, including the retrograde flow $\delta {\boldsymbol v}$.
They saturate at finite amplitudes, which will be determined analytically and numerically in  sections \ref{sec_nonlinear_analysis_supplement} and \ref{sec_numerics_supplement}.

\section{Linear stability analysis for higher order harmonics}
\label{sec_linear_stability_supplement}
The  critical polymerization velocity $v_p^c$ (\ref{vpc}), for which the spontaneous symmetry breaking (cell polarization) sets in, is determined here by a more detailed linear stability analysis of the homogeneous concentration $c_0$ of actin filaments in the cell cortex. Assuming the origin of the coordinate system at the cell center, the position of the cortex along the cell boundary is given by 
\begin{equation}\label{eq_cell_surface_descr}
\bm r (\theta, \varphi)=  R_0 \left[ 1 + \varrho(\theta) \right] \bm e_r,
\end{equation}
with the radial unit vector $\bm e_r = (\sin \theta \cos \varphi, \sin \theta \sin \varphi, \cos \theta)$.
$R_0=[A_0/(4\pi)]^{1/2}$ is the radius of the undeformed spherical cell and the shape function $\varrho(\theta)$ describes deviations from the spherical shape. 
For simplicity, we assume axial symmetry around the $z$-axis.
The concentration of actin filaments is expanded into a series of spherical harmonics,
\begin{equation}
\label{c_expansion}
c (\bm r, t) = c_0 + \delta c (\bm r,t),
\end{equation}
with
\begin{equation}
	\delta c (\bm r,t) = \sum_{l=1}^{\infty}  \delta \hat c_{l}(t) Y_{l} (\theta).
\end{equation}
Here, $\delta \hat{c}_{l}(t)$ is the time-dependent amplitude of the $l$-th harmonic and
\begin{equation}\label{eq_def_scalar_SH}
	Y_{l} (\theta) = \sqrt{\frac{2l+1}{4 \pi} } P_l (\cos \theta)
\end{equation}
is the axisymmetric scalar spherical harmonic where $P_l (x)$ is the Legendre polynomial of degree $l$. The first Legendre polynomials are given by
\begin{equation}
\begin{aligned}
	P_0 (\cos \theta) &= 1, \\
	P_1 (\cos \theta) &= \cos \theta, \\
	P_2 (\cos \theta) &= \frac{1}{2} \left( 3 \cos^2 \theta -1 \right).
\end{aligned}
\end{equation}
The linear stability analysis for arbitrary $l$ follows the procedure used for $l=1$ above.
The only difference is that the tension force in the force balance equation {(see main text)} is not zero for $l>1$.

A straightforward calculation yields the leading-order evolution equation for $\delta \hat c_l$ as
\begin{equation}
	\label{eq_c_evolution_SI}
	\delta \dot{\hat c}_l = \lambda_l \delta \hat c_l
\end{equation}
where the growths rates $\lambda_l$ are obtained as
\begin{equation}
    \label{lambdal}
	\lambda_l = \frac{2l(l+1)(\eta_s+\eta_b)}{(2\eta_s+\eta_b)l(l+1)-2\eta_s}\frac{v^0_p c_0}{R_0 c_r} e^{-\frac{c_0}{c_r}}- l(l+1) \frac{D}{R_0^2}  - \beta.
\end{equation}
Note that the first term in (\ref{lambdal}) decreases with increasing $l$ (excluding the $l=0$ case).
This means that the loss of stability of the unpolarized solution always occurs via the instability of the first harmonic.
Equation (\ref{lambdal}) has a particularly simple expression for $l=1$, see Eq.\ \eqref{lambda1}.

The instability takes place for $\lambda_1 >0$ which corresponds to the case of $v_p^0 > v_p^c$, with the critical polymerization velocity
\begin{equation}
\label{polycrit_1}
v_p^c = \frac{c_r}{2c_0} \left(\frac{2D}{R_0} + \beta R_0 \right) e^\frac{c_0}{c_r}.
\end{equation}

\section{Amplitude of the concentration modulation closely above threshold}
\label{sec_nonlinear_analysis_supplement}
In the following, the weakly non-linear analysis is presented which allows us to determine analytically the bifurcation type for the instability of the concentration of actin filaments due to polymerization.
The general procedure is to solve the cortex force balance equation under the condition that the cell shape remains fixed in the reference frame co-moving with the cell. This yields the cortex velocity which together with the polymerization velocity determines the full velocity. The latter is then used to solve the advection-diffusion equation on the co-moving cell surface to obtain expressions for the time-evolution of the actin concentration.\\

Assuming small deviations from a spherical cell shape, the shape function in Eq.\ \eqref{eq_cell_surface_descr} can then be expanded into Legendre Polynomials according to
\begin{equation}\label{eq_rho_expansion}
	\varrho (\theta) = \sum_{l=2}^{\infty} \varrho_l P_l (\cos \theta),
\end{equation}
where $\varrho_l$ are the shape coefficients. Note that the Legendre polynomials correspond up to prefactors to the scalar spherical harmonics as used in Eq.\ \eqref{eq_def_scalar_SH}.
Eq.\ \eqref{eq_rho_expansion} describes cell deformations where the lowest non-vanishing order is $l=2$. We have $\varrho_0 = 0$ due to the conservation of the membrane area and $\varrho_1=0$ since it describes a translation of the cell. 
Forces and velocities are expanded into vector spherical harmonics $\bm Y_{j,l} (\theta, \varphi)$ which are given by
\begin{equation}
\begin{aligned}\label{eq_vector_spherical_harmonics}
\boldsymbol{Y}_{1,l} (\theta, \varphi) &=  \left[ {\fat{\nabla}}^s - (l+1) \bm e_r \right] Y_{l} (\theta)\,, \\
\boldsymbol{Y}_{2,l} (\theta, \varphi) &=  \left[ {\fat{\nabla}}^s + l \bm e_r \right] Y_{l} (\theta)\,, \\
\boldsymbol{Y}_{3,l} (\theta, \varphi) &=  \bm e_r \times {\fat{\nabla}}^s Y_{l} (\theta)\,.
\end{aligned}
\end{equation}
With this, the cortex velocity can be written as
\begin{equation}\label{eq_expand_vcortex_axissymmetric_VSH}
\fat{v}_c (\theta, \varphi) = \sum_{j = 1}^{3} \sum_{l = 0}^{\infty} v^c_{j,l} \fat{Y}_{j,l} (\theta, \varphi),
\end{equation}
and the passive viscous force as
\begin{equation}\label{eq_expand_fvisc_axissymmetric_VSH}
\fat{f}_v (\theta, \varphi) = \sum_{j = 1}^{3} \sum_{l = 0}^{\infty} f^v_{j,l} \fat{Y}_{j,l} (\theta, \varphi),
\end{equation}
where $v^c_{j,l}$ and $f^v_{j,l}$ are constant coefficients.
The actin filament concentration in the cortex is expanded up to second order,
\begin{equation}\label{eq_init_conc_second_order}
c (\bm r, t) = c_0 + \varepsilon \delta \hat c_1(t) P_1 (\cos \theta) + \varepsilon^2 \delta \hat c_2(t) P_2 (\cos \theta),
\end{equation}
where $\varepsilon > 0$ is a small parameter and $\delta \hat c_1$ and $\delta \hat c_2$ are the amplitudes of the first and second harmonic of the concentration, respectively.
{From the expression for $\boldsymbol{v}_p (c)$ (see main text)}
one obtains with equation \eqref{eq_init_conc_second_order} the polymerization velocity
\begin{equation}\label{eq_actin_vp_higher_order}
\bm v_p (\theta, \varphi) = v^0_p e^{-\bar{c}_0} \Biggl\{ 1 - \varepsilon \delta \bar{c}_1 \cos \theta + \frac{\varepsilon^2}{2}  \left[ \delta \bar{c}_2 + \cos^2 \theta \left( \delta \bar{c}_1^2 - 3 \delta \bar{c}_2 \right) \right] - \frac{\varepsilon^3}{6} \delta \bar{c}_1 \cos \theta \left[ \left( \delta \bar{c}_1^2 - 9 \delta \bar{c}_2 \right) \cos^2 \theta + 3 \delta \bar{c}_2 \right] \Biggr\} \bm e_r,
\end{equation}
where the dimensionless concentrations $\bar{c}_0 := \frac{c_0}{c_r}$ and $\delta \bar{c}_l := \frac{\delta \hat c_l}{c_r}$ have been introduced.
The assumption of a fixed shape implies that the normal component of the full velocity in the co-moving reference frame has to vanish,
\begin{equation}\label{eq_fixed_shape_condition}
\fat{v} (\theta, \varphi) \bm \cdot \bm e_r = 0,
\end{equation}
where
\begin{equation}\label{eq_full_velocity}
	\fat{v} (\theta, \varphi) = \bm v_c (\theta, \varphi) + \bm v_p (\theta, \varphi).
\end{equation} 
The $O(\varepsilon)$ contribution to the cortex velocity is written as,
\begin{equation}\label{eq_v0_epsilon}
\fat{v}_c = \varepsilon v_p^0 e^{- \bar{c}_0} \delta \bar{c}_1 \fat{e}_z,
\end{equation}
as described in the main text.
For the detailed expressions for the coefficients of the passive viscous force $\bm f_v$ and the tension force $\bm f_t$ we refer to the supplementary information of Ref.\ \cite{Farutin2019}. Solving the force balance {(see main text)} together with the fixed shape condition \eqref{eq_fixed_shape_condition} for $l=2$ then yields the shape coefficient
\begin{equation}\label{eq_rho_2}
\varrho_2 = 4 \varepsilon^2 v_p^0 \frac{  \eta_s (\eta_s + \eta_b) e^{- \bar{c}_0} \left( \delta \bar{c}_1^2 - 3 \delta \bar{c}_2 \right) }{3 \Delta P R_0^2 \left( 5 \eta_s + 3\eta_b \right)},
\end{equation}
and with that the coefficients for the cortex velocity in $O(\varepsilon^2)$ order,
\begin{equation}\label{eq_v_cortex_coeffs_second_order}
\begin{aligned}
v^c_{1,2} &= 2 \varepsilon^2 v_p^0  \frac{\sqrt{5 \pi} (3 \eta_s + \eta_b) }{75 (5 \eta_s + 3 \eta_b)} R_0 e^{- \bar{c}_0} \left( \delta \bar{c}_1^2 - 3 \delta \bar{c}_2 \right) \\
v^c_{2,2} &= -4 \varepsilon^2 v_p^0  \frac{\sqrt{5 \pi} (4 \eta_s + 3 \eta_b) }{75 (5 \eta_s + 3 \eta_b)} R_0 e^{- \bar{c}_0} \left( \delta \bar{c}_1^2 - 3 \delta \bar{c}_2 \right) \\
v^c_{3,2} &= 0.
\end{aligned}
\end{equation}
Together with Eq.\ \eqref{eq_actin_vp_higher_order} and Eq. \eqref{eq_full_velocity}, the full velocity $\bm v (\theta, \varphi)$ is herewith expressed as a function of the concentration harmonics $\delta \bar{c}_1$ and $\delta \bar{c}_2$.
The tangential retrograde flow, the full velocity in the co-moving frame, is then given by
\begin{equation}\label{eq_full_velocity_comoving}
	\fat{v} (\theta, \varphi) = - \varepsilon v_p^0 e^{-\bar{c}_0} \delta \bar{c}_1  \sin \theta \bm t^\theta + \varepsilon^2 v_p^0 e^{-\bar{c}_0} \frac{\eta_s + \eta_b}{5 \eta_s + 3 \eta_b} \left( \delta \bar{c}_1^2 - 3 \delta \bar{c}_2 \right) \cos \theta \sin \theta \bm t^\theta.
\end{equation}
The retrograde flow can be written as the surface gradient of a flow potential $\psi (\theta)$, 
\begin{equation}\label{eq_def_flow_potential}
\fat{v} (\theta, \varphi) = \bm \nabla^s \psi (\theta),
\end{equation}
yielding upon integration
\begin{equation}\label{eq_flow_potential_perturbation}
\psi  = 3 \varepsilon \cos \theta \frac{ v_p^0 R e^{-\bar{c}_0 }}{(30 \eta_b + 50 \eta_s)} \Biggl\{ - \frac{5}{3} \varepsilon \cos \theta (\eta_b + \eta_s) \left( \delta \bar{c}_1^2 - 3 \delta \bar{c}_2 \right) + \delta \bar{c}_1 \left( \eta_b + \frac{5}{3} \eta_s \right) \left[ \varepsilon^2 \left( \delta \bar{c}_1^2 - 4 \delta \bar{c}_2 \right) + 10 \right] \Biggr\}.
\end{equation}
The advection term in equation (\ref{cdot_advective})  can be expressed in terms of the flow potential according to
\begin{equation}\label{eq_advection_term_flow_potential}
\bm \nabla^s \bm \cdot  \left( \fat{v} c \right)  = \frac{1}{2} \left[ \Delta^s \left( c  \psi  \right) + c  \Delta^s \psi  - \psi  \Delta^s c  \right].
\end{equation}
With this, $\dot{c}$ can be calculated from equation (\ref{cdot_advective}). Projecting the result on the respective Legendre polynomials yields 

\begin{align}
\label{eq_coeff_cdot_actin_non_linear_1}
\notag
 \delta \dot{\bar{c}}_{1} &= \varepsilon \delta \bar{c}_1 \left( 2 \bar{v}_p^0 \bar{c}_0 - 2 \bar{D} - 1 \right)\\ 
&+  \frac{\varepsilon^3 \bar{v}_p^0 \delta \bar{c}_1}{5 \left(3 \eta_b + 5 \eta_s \right)}  \biggl\{ \left[ -2 \left( \eta_b + \eta_s \right) + \bar{c}_0 \left( 3 \eta_b + 5 \eta_s \right) \right] \delta \bar{c}_1^2 - 4 \delta \bar{c}_2 \left[ \eta_s + \bar{c}_0 \left( 3 \eta_b + 5 \eta_s \right) \right] \biggr\} + \mathcal{O} (\varepsilon^5), \\
\label{eq_coeff_cdot_actin_non_linear_2}
\notag
 \delta \dot{\bar{c}}_{2} &=  \frac{- \varepsilon^2}{\left( 3 \eta_b + 5 \eta_s \right)} \biggl\{ 2 \bar{v}_p^0 \left[ \left( \eta_s + \eta_b \right) \bar{c}_0 \delta \bar{c}_1^2 - 3 \eta_b \left( \bar{c}_0 \delta \bar{c}_2 + \delta \bar{c}_1^2 \right) - \eta_s \left( 3 \bar{c}_0 \delta \bar{c}_2 + 5 \delta \bar{c}_1^2 \right) \right] \biggr. \\
&\biggl. + \left( 3 \eta_b + 5 \eta_s \right) \delta \bar{c}_2 \left( 6 \bar{D} + 1 \right) \biggr\} + \mathcal{O} (\varepsilon^4),
\end{align}
where for simplicity the dimensionless parameters
\begin{equation}
\label{eq_dimless_params_2}
\delta \dot{\bar{c}}_l := \frac{\delta \dot{\hat c}_l}{c_r \beta}, \qquad \bar{v}_p^0 := \frac{v_p^0}{R_0 \beta} e^{-\bar{c}_0}, \qquad \bar{D} := \frac{D}{R_0^2 \beta}, \qquad \bar{\eta} := \frac{\eta_s}{\eta_b}
\end{equation}
have been introduced.
Making use of the adiabatic approximation, $\delta \dot{\bar{c}}_{2} = 0$,
one obtains from equation \eqref{eq_coeff_cdot_actin_non_linear_2} an expression for $\delta \bar{c}_2$, which upon substitution in equation \eqref{eq_coeff_cdot_actin_non_linear_1} yields an expression for the time-evolution of the first harmonic,
\begin{equation}\label{eq_first_harmonic_third_order}
\delta \dot{\bar{c}}_{1} = ( \bar{\lambda}_1   + \nu \delta \bar{c}_1^2) \delta \bar{c}_1.
\end{equation}
The prefactors $\bar{\lambda}_1$ and $\nu$ are given by
\begin{equation}
\bar{\lambda}_1 = 2 \bar{c}_0 ( \bar{v}_{p}^0 - \bar{v}_{p}^c),
\end{equation}
where 
\begin{equation}\label{eq_vpoly_crit}
\bar{v}_{p}^c = \frac{2\bar{D}+1}{2 \bar{c}_0}
\end{equation}
is the critical polymerization velocity
and
\begin{equation}\label{eq_non_linear_coefficient}
\nu = \frac{\left(2 \bar{D} + 1 \right) \biggl\{ \left[ \left( \frac{3}{2} + 8 \bar{D} \right) \bar{\eta} + \left( 5 \bar{D} + 1 \right) \right] \bar{c}_0^2 - \left[ \left( 5 + 12 \bar{D} \right) \bar{\eta} + \left( 3 + 8 \bar{D} \right) \right] \bar{c}_0 - \left(1 + 2 \bar{D} \right) \bar{\eta} \biggr\} }{5 \bar{c}_0^2 \left[ \left( 1 + 12 \bar{D} \right) \bar{\eta} + 6 \bar{D} \right]}.
\end{equation}
With  $\lambda_1 = \beta \bar{\lambda}_1$, equation \eqref{eq_first_harmonic_third_order} corresponds to equation (4) of the main text, and equation \eqref{eq_vpoly_crit} in its dimensional form to equation (\ref{vpc}).
The fixed points of equation \eqref{eq_first_harmonic_third_order}, as discussed in the main text, are obtained by setting $\delta \dot{\bar{c}}_1 = 0$. The migration speed $v_0 = |\bm v_0|$ as plotted in Fig.\ (3) of the main text is then obtained from equation \eqref{eq_v0_epsilon} by substituting the stationary solution for $\delta \bar c_1$.

\begin{figure}[h]\centering
	\includegraphics{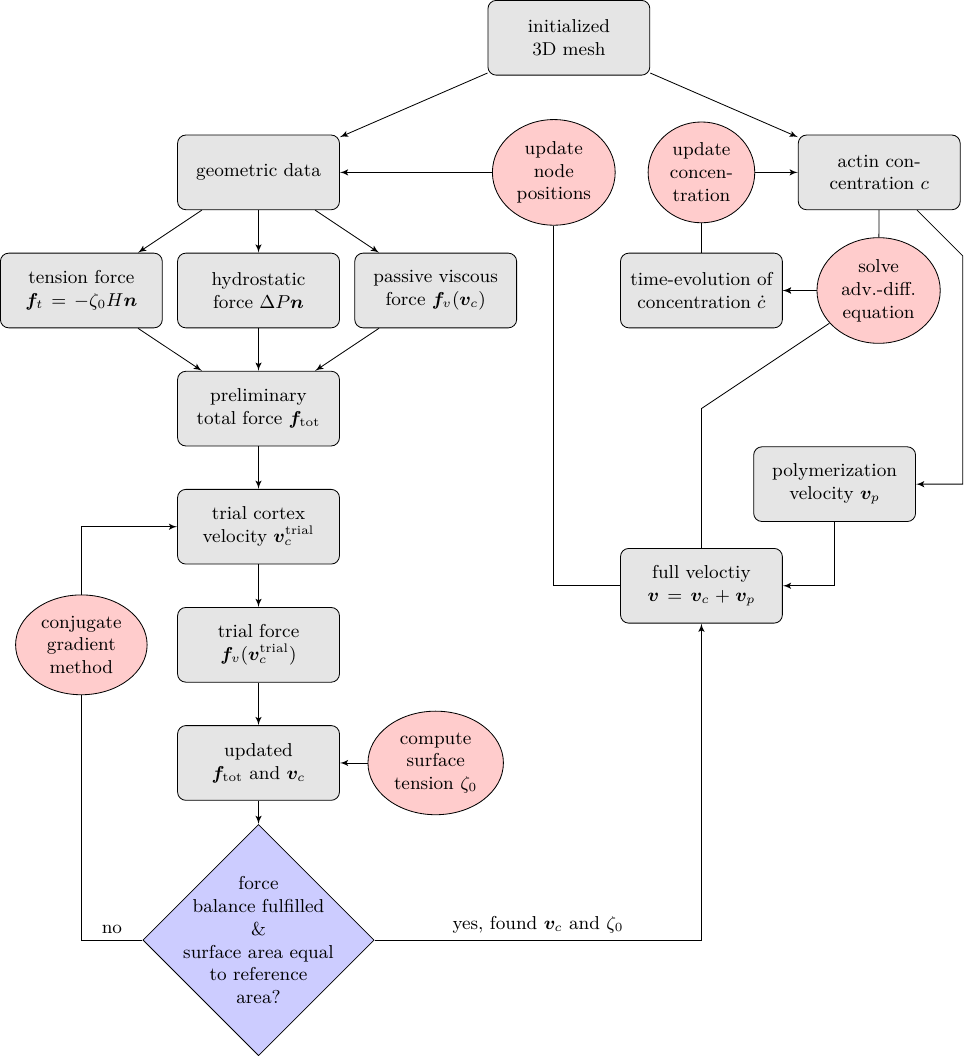}
	\caption{Flow chart of the numerical algorithm. Gray boxed symbolize sets of data, arrows indicate the order of data processing, and pink ellipses show the employed methods and procedures.}
	\label{fig_flow_chart_numerical_procedure}
\end{figure}
\section{Numerical procedure}
\label{sec_numerics_supplement}
In the following, the algorithm for the numerical solution of the thin shell model as described in the main text is presented.
The cortex and the membrane form together the cell boundary.
It is represented by a 3D mesh that consists of $N=642$ points (nodes) that form triangles (faces), where each node $a$ is characterized by its position $\boldsymbol r^a$.
Similarly, the cortex velocity $\boldsymbol v_c^a$, the polymerization velocity $\boldsymbol v_p^a$ and the total velocity $\boldsymbol v^a$ are tracked for each node $a$.
The initial cell shape is set to a sphere with a reference surface area $A_0 = 4 \pi R_0^2$.

At each iteration, we compute the cortex velocity $\boldsymbol v_c$ at each mesh node by solving the force-balance equation
\begin{equation}
\label{forcebalanceN}
\boldsymbol f_v(\boldsymbol v_c)+\boldsymbol f_e= \boldsymbol 0,
\end{equation}
where $\boldsymbol f_v$ is the viscous force and $\boldsymbol f_e$ are the conservative forces in the problem, such as the membrane tension forces and the osmotic pressure difference.

Equation (\ref{forcebalanceN}) is solved numerically using weak formulation:
The viscous forces are related to the dissipation function $\mathcal D$:
\begin{equation}
\label{varD}
\delta \mathcal D=-\oint \delta \boldsymbol v_c\boldsymbol\cdot\boldsymbol f_v(\boldsymbol v_c) d^2r,
\end{equation}
where the integration is taken along the cell surface and
\begin{equation}\label{dissipation}
\mathcal D= \oint \left[ \frac{\eta_b}{8} \left( \dot{g}_{\alpha \beta} g^{\alpha \beta} \right)^2 + \frac{\eta_s}{4} \left( g^{\beta \gamma} \dot{g}_{\alpha \beta} \dot{g}_{\gamma \delta} g^{\alpha \delta} \right) \right] d^2r.
\end{equation}
Here $g_{\alpha\beta}=\partial_\alpha\boldsymbol r\boldsymbol\cdot\partial_\beta\boldsymbol r$ is the metric tensor and $g^{\alpha\beta}$ is its inverse.
The time derivative of $g_{\alpha\beta}$ is related to the surface strain tensor $\mathsf E^s=\boldsymbol\nabla^s\otimes\boldsymbol v^c\boldsymbol\cdot\mathsf I^s+\mathsf I^s\boldsymbol\cdot(\boldsymbol\nabla^s\otimes \boldsymbol v^c)^T$ as 
\begin{equation}
\label{metric}
\partial_\alpha r_ig^{\alpha\beta}\dot g_{\beta\gamma}g^{\gamma\delta}\partial_\delta r_j=\partial_\alpha r_ig^{\alpha\beta}(\partial_\beta v^c_k\partial_\gamma r_k+\partial_\beta r_k\partial_\gamma v_k) g^{\gamma\delta}\partial_\delta r_j=E^s_{ij}
\end{equation}
The conservative forces are related to the corresponding energy function $\mathcal E$
\begin{equation}
\label{varE}
\delta \mathcal E=-\oint \delta \boldsymbol r\boldsymbol\cdot\boldsymbol f_e d^2r,
\end{equation}
\begin{equation}\label{energy}
\mathcal E=\zeta_0\oint d^2r+\Delta P\oint (\boldsymbol r\boldsymbol\cdot\boldsymbol n) d^2r.
\end{equation}
We express $\mathcal D$ and $\mathcal E$ as functions of positions and velocities of the mesh nodes, where the metric tensor is computed on each mesh triangle as discussed, for example, in \cite{Farutin2014}.

Equation (\ref{forcebalanceN}) is discretized as
\begin{equation}
\label{forcebalance2}
\frac{\partial\mathcal D}{\partial \boldsymbol v_c^a}+\frac{\partial\mathcal E}{\partial \boldsymbol r^a}= \boldsymbol 0,
\end{equation}
for all $a = 1,...,N$.
Equations (\ref{forcebalance2}) is a system of linear equations, which allow us to compute $\boldsymbol v_c^a$ for all $a$ for given positions of the mesh points and values of $\zeta_0$ and $\Delta P$.
In this case, the value of $\zeta_0$ is not known, and instead we impose the fixed-area condition $A=A_0$, which translates into a linear constraint on the velocities $\boldsymbol v^a$ \cite{Farutin2014}.
Because the Eqs. (\ref{forcebalance2}) are obtained as a gradient of a convex quadratic form of the velocities of the mesh nodes, they, together with the linear constraint of area conservation, can be solved efficiently by a conjugate gradient method.

Fig.\ \ref{fig_flow_chart_numerical_procedure} shows the numerical procedure of our algorithm.  At the beginning of the simulation, the parameters are defined and the mesh initialized. The initial concentration of actin filaments in the cortex at node $a$ is given by $c(0) = c_0 + \delta c_1 (\bm r^a - \bm r_{cm}) \bm \cdot \bm e_z /R_0$, where $\delta c_1$ is a perturbation of the homogeneous concentration $c_0$ and $\bm r_{cm}$ is the position of the cell's center of mass. That is, a small perturbation $\delta c_1 z$ along the $z$-axis is added to the homogeneous concentration $c_0$.
Entering the time-loop, first a set of instantaneous geometric data is calculated from the node positions. This includes the local normal vectors and areas of the triangles, as well as the global surface area $A_s$ and volume $V$ of the cell. From this, the variations of $\mathcal D$ and $\mathcal E$ are computed.
These two contributions are summed up to a residue of Eq. (\ref{forcebalance2}), which is, in general, non-zero since the initial assumptions for the surface tension $\zeta_0$ and cortex velocity $\bm v_c$ do not correspond to the correct values.\\

The surface tension and cortex velocity field are obtained via the conjugate gradient method under the constraint that the surface area of the cell is equal to the initial reference area.
Continuing, we can compute the full velocity $\bm v$ from the polymerization velocity $\bm v_p$ which is calculated from the actin concentration $c$.
$\bm v$ is then used together with $c$ to solve the advection-diffusion equation (\ref{cdot_advective}) which yields $\dot{c}$.
The concentration is then updated according to $c(\bm r, t + \Delta t) = c(\bm r, t) + \dot{c} (\bm r, t) \Delta t$, where $\Delta t$ is the time-step. Finally, the node positions are updated as $\bm r(t + \Delta t) = \bm r(t) + \bm v (\bm r, t) \Delta t$ which concludes the time-loop.\\

Two additional procedures are required to ensure the numerical stability of the method:
First, since the dissipation function is invariant upon addition of an arbitrary rotation or translation to the velocity field, the solution of (\ref{forcebalance2}), just as the solution of the original problem, is not unique.
We resolve this ambiguity by choosing the solution with zero angular and translational velocities.
Second, the advection of the mesh nodes by the retrograde flow distorts the mesh triangles.
This distortion is countered by moving the mesh nodes along the cell surface, as discussed in \cite{Farutin2014} and computing the filament concentration at the new positions by interpolation.
\section{Parameters}
\label{sec_parameters}
The parameters listed in table \ref{tab_params} are used to obtain the numerical and analytical results in the main text, if not mentioned otherwise. The table also shows typical values for the parameters in physical units from the literature.
For the numerical simulations a time step of $\Delta t = 10^{-5}$ is used. For the numerically obtained supercritical bifurcation (see Fig. 3 A in the main text) and the lower branch of the subcritical bifurcation (see Fig. 3 B in the main text) an initial perturbation of the homeostatic actin concentration of $\delta c_1 = 0.03$ is chosen, while the upper branch of the subcritical bifurcation is obtained for $\delta c_1 = 3$. 
\begin{table}[h]
	\begin{tabular}{ c c c c }
		\hline
		parameter name & symbol & simulation & typical value  \\ 
		& &  (arbitrary units) & (physical units) \\
		\hline \hline
		cell radius & $R_0$ & $1$ & $\SI{10}{\micro \meter}$ \\
		pressure drop across membrane & $\Delta P$ & $300$ & $\SIrange[range-units = brackets, range-phrase = -]{300}{700}{\pascal}$ \cite{doi:10.1126/science.1256965} \\ 
		cortex shear viscosity & $\eta_s$ & $1$ & $\SI{2.7}{\milli \pascal \second \meter}$ \cite{Bergert2015} \\ 
		cortex bulk viscosity & $\eta_b$ & $1$ & {-} \\ 
		homeostatic F-actin concentration & $c_0$ & $3$ & {$\SIrange[range-units = brackets, range-phrase = -]{100}{1400}{\per \micro \meter \squared}$} \cite{ABRAHAM19991721, 10.1371/journal.pone.0004810, RAZBENAROUSH20172963}  \\ 
		{cortex thickness} & {$h$} & {-} & {$\SI{190}{\nano \meter}$ \cite{clark2013monitoring}} \\
		{G-actin volume concentration} & {$\mu$} & {-} & {$\SIrange[range-units = brackets, range-phrase = -]{8.5}{1200}{\micro \mol \per \liter}$ \cite{ABRAHAM19991721, RAZBENAROUSH20172963}} \\
		reference actin concentration & $c_r$ & $1$ & - \\ 
		F-actin turnover rate & $\beta$ & $1$ & $\SI{0.1}{\per \second}$ \cite{mogilner2009mathematics}\\ 
		F-actin diffusion coefficient & $D$ & $0.03$ & $\SI{0.03}{\micro \meter \squared \per \second}$ \cite{RAZBENAROUSH20172963, Copos} \\ 
		F-actin polymerization speed & $v_p^0$ & $3.8$ & $\SIrange[exponent-to-prefix, range-units = brackets, range-phrase = -]{5e-3}{3}{\micro \meter \per \second}$ \cite{10.1083/jcb.200806185, doi:10.1146/annurev.biophys.29.1.545, doi:10.1126/science.1100533} \\ 
		\hline
		non-dimensional homeostatic F-actin conc. \ & $\bar{c}_0 = c_0 / c_r$ &  $3$ & \\
		non-dimensional F-actin diffusion coeff. & $\bar{D} = D / (R_0^2 \beta)$ & $0.03$ & \\
		non-dimensional cortex viscosity ratio & $\bar{\eta} = \eta_s / \eta_b$ & $1$ & \\
		non-dimensional F-actin polym.\ speed & $\bar{v}_p^0 = v_p^0 e^{-c_0/c_r} / (R_0 \beta)$ \ & $0.19$ & \\
		\hline
	\end{tabular}
	\caption{Parameters used for simulations and analytical considerations. Third column at the top shows parameters given in simulation units, the resulting non-dimensional parameters marked with bars are displayed at the bottom. Last column at the top shows typical ranges of physical parameters from the literature.}
	\label{tab_params}
\end{table}

{We assume that the shear and bulk 2D viscosities of the cortex are of similar order of magnitude \cite{Bergert2015}, since filaments are allowed to move freely along the direction perpendicular to the membrane. For simplicity, we set $\eta_s = \eta_b$ throughout our study.}

The order of magnitude of the retrograde flow can be estimated from available literature parameters.
Let us assume a spherical cell with radius  $R_0 = \SI{10}{\micro \meter}$, 
a cortical viscosity of {$\eta_s = \eta_b = \SI{2.7}{\milli \pascal \second \meter}$} \cite{Bergert2015},
a rate at which F-actin stops treadmilling (due to the attachment of a capping protein) of $\beta = \SI{0.1}{\per \second}$ \cite{mogilner2009mathematics},
{an actin surface diffusion coefficient of $D =\SI{0.03}{\micro \meter \squared \per \second}$ \cite{RAZBENAROUSH20172963, Copos},
and a surface concentration of polymerizing actin filaments $c_0 = \SI{1370}{\per \micro \meter \squared}$ \cite{ABRAHAM19991721}.
Following our estimate in section \ref{sec_estimate_monomer_competition}, we set $c_r = \mu h$. 
Herein, $\mu = \SI{8.5}{\micro \mol \per \liter}$ \cite{RAZBENAROUSH20172963} is the bulk concentration of G-actin monomers available for polymerization and $h = \SI{190}{\nano \meter}$ a length scale at the order of magnitude of the cortex thickness.}
With this we find analytically a supercritical bifurcation with a critical polymerization speed of 
{$v_p^c = \SI{1.5}{\micro \meter \per \second}$}. 
Assuming $v_p^0 = \SI{1.7}{\micro \meter \per \second}$ \cite{10.1083/jcb.200806185, doi:10.1146/annurev.biophys.29.1.545}, we obtain a maximum retrograde flow speed of 
{$\SI{0.17}{\micro \meter \per \second}$}.
This value agrees well with typical cell migration speeds measured in experiments \cite{Ruprecht_2015_CCT}.

{ 
\section{Estimate for the competition of polymerizing actin filaments for monomers}
\label{sec_estimate_monomer_competition}
Here we give a justification for the anti-correlational relation between the polymerization velocity and the cortical F-actin concentration, as assumed in our model.
There is experimental evidence that the growth of actin filaments is slowed by a competition of polymerizing filaments for available G-actin monomers. 
In Ref.\ \cite{colin2023recycling}, actin polymerizes against a polystyrene bead in a microwell which contains a limited amount of components.
The growth velocity $v_p$ of the comet, that is, the polymerization velocity, is observed to decay exponentially with time $t$. 
In the following we denote the monomer concentration, i.e., the number of polymerizable G-actin monomers per unit volume, as $\mu$.
By fitting the function
\begin{equation}\label{eq_vp_lit}
	v_p (t) = B e^{- C t}
\end{equation}
to the data in Ref.\ \cite{colin2023recycling}, we extract the parameters $B = \SI{1.184}{\micro \meter \per \minute}$ and $C = \SI{1.248}{\per \hour}$ for an initial monomer concentration of $\mu^0 = \SI{3}{\micro \mol \per \liter}$.
Assuming that $\mu$ in the microwell is decreased by the polymerization of filaments, we write
\begin{equation}\label{eq_cmono_lit}
	\mu (t) = \mu^0 - \gamma \int_{0}^{t} v_p(t') dt',
\end{equation}
where $\mu^0$ is the initial monomer concentration in the microwell  and $\gamma > 0$ is a constant.
Using equation \eqref{eq_vp_lit} and assuming that for large times all monomers in the system are depleted, i.e., $\lim_{t \to \infty} \mu(t) = 0$, one obtains $\gamma = \mu^0 C / B$
and equation \eqref{eq_cmono_lit} yields
\begin{equation}\label{eq_cmono_lit2}
	\mu (t) = \mu^0  e^{-C t}.
\end{equation}
From equation \eqref{eq_vp_lit} and \eqref{eq_cmono_lit2} then follows a linear relation between the polymerization velocity and monomer concentration,
\begin{equation}\label{eq_vp_monomer}
	v_p (\mu) = \frac{B }{\mu^0} \mu.
\end{equation}
Ref. \cite{colin2023recycling} suggests that the polymerization speed is limited by competition for monomers if the number of available monomers is sufficiently small compared to the number of polymerizing filaments.
{We define the dimensionless ratio of monomer volume density to filament surface density according to
\begin{align}
	\bar{\mu} = \frac{\mu h}{c}.
\end{align}
Herein, $h$ is a characteristic length scale which we equate with the thickness of the cortex. 
This reflects our assumption that the polymerization rate of cortical filaments is fast and therefore limited by monomer diffusion which brings new building blocks from a bulk reservoir in the cytoplasm to the outer edge of the cortex.}
With this we write the polymerization speed up to first order as
{
\begin{equation}\label{eqAnsatzVPolyRatio}
	v_p \left( \bar{\mu}\right) = V_0 + V_1 \bar{\mu} + \mathcal{O} \left( \bar{\mu}^2 \right)
\end{equation}
}
with constants $V_0$ and $V_1$.
Comparing equation \eqref{eqAnsatzVPolyRatio} to equation \eqref{eq_vp_monomer} yields
\begin{equation}\label{eqV1}
	V_0 = 0, \quad {V_1 = \frac{B c}{\mu^0 h}}.
\end{equation}
In the \textit{in vitro} study \cite{colin2023recycling}, the filament concentration $c$ is constant with $6700$ filaments at every cross-section $A$ of the actin tail where $A = \SI{12}{\micro \meter \squared}$, yielding $c = \SI{560}{\per \micro \meter \squared}$.
{Assuming $h = \SI{190}{\nano \meter}$ \cite{clark2013monitoring}}, with this we obtain from equation \eqref{eqV1}

\begin{equation}\label{eqV1Numbers}
	{V_1 =  \frac{\SI{1.184}{\micro \meter \per \minute} \times \SI{560}{\per \micro \meter \squared}}{\SI{3}{\micro \mol \per \liter} \times \SI{190}{\nano \meter}} = \SI{32.2}{\nano \meter \per \second}.}
\end{equation}
With this value for $V_1$ one can compute from equation \eqref{eqAnsatzVPolyRatio} the polymerization velocity for a given ratio of monomer to filament concentration in the regime where filament growth is limited by competition for monomers.
The ratio between the monomer and filament concentration in the \textit{in vitro} study \cite{colin2023recycling} is approximately given by
\begin{equation}\label{eqRatioInVitro}
	{\bar{\mu}^\text{vitro} = \frac{\SI{3}{\micro \mol \per \liter} \times \SI{190}{\nano \meter}}{\SI{560}{\per \micro \meter \squared} } = 0.61.}
\end{equation}
The  values for the concentrations \textit{in vivo} are available in the literature, e.g., for migrating fibroblasts  where $\mu^\text{vivo} = \SI{8.5}{\micro \mol \per \liter}$ and $c^\text{vivo} = \SI{1370}{\per \micro \meter \squared}$ have been reported \cite{ABRAHAM19991721}. 
With this we find
\begin{equation}\label{eqRatioInVivo}
	{\bar{\mu}^\text{vivo} = 0.71},
\end{equation}
which is comparable to the ratio in the \textit{in vitro} system, see equation \eqref{eqRatioInVitro}.
This shows that even for physiological filament and monomer concentrations, the filament growth can be potentially limited by a competition of monomers.
Note that Eq.\ \eqref{eqAnsatzVPolyRatio} suggests a relation  $v_p \propto  c^{-1}$, while in our model we assume that the polymerization velocity decreases exponentially with the actin concentration. 
{Below in section \ref{sec_weakly_nonlinear_inverse} we compare the migration speed obtained by a weakly nonlinear analysis for both polymerization laws and find qualitatively similar behavior.}
}

{ 
\section{Case of zero restoration rate}
The restoration rate $\beta$ which accounts for depolymerization counteracts the instability described by us. This is evident from Equation (3) of the main text where the critical polymerization rate grows with increasing $\beta$.
In Figure \ref{fig_beta_zero} we plot the retrograde flow speed as a function of the polymerization strength for $\beta = 0$ (no depolymerization) with parameters otherwise equal to the ones used for Figure 3A in the main text. 
Compared to the case of $\beta = 1$, the bifurcation changes from supercritical to subcritical, which is in agreement with our analytical solution. The analytically obtained critical polymerization velocity also agrees well with the numerical onset of retrograde flow. 

\begin{figure}[h]\centering
	\includegraphics{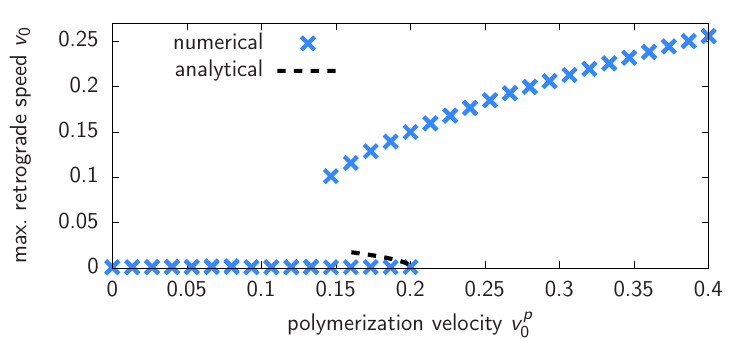}
	\caption{{  Maximum retrograde flow speed $v_0$ as a function of polymerization strength $v_0^p$ in the case of zero restoration ($\beta = 0$). Numerical results are shown as blue crosses and the analytical result as the black dashed line.}}
	\label{fig_beta_zero}
\end{figure}
}

{
\section{Linear stability analysis including cortical resistance to compression}
For simplicity, we neglected in our model the elastic resistance of the cortex.
In the following we perform a linear stability analysis of an extended model which accounts for such an effect. We derive a modified growth rate and critical polymerization speed, showing that resistance to compression counter-acts the described instability.
By a comparison to physical parameters from the literature we show that our described instability can take place even if cortical resistance to compression is taken into account.

Following previous work \cite{Farutin2019}, we include cortical resistance to compression by an additional term which depends linearly on the actin concentration in the passive surface stress in Equation \eqref{passive},
\begin{equation}
	\label{surfaceStressAlpha}
	\sigma^s=\eta_b(\boldsymbol\nabla^s\boldsymbol\cdot\boldsymbol v_c)\mathsf I^s+\eta_s[\boldsymbol\nabla^s\otimes\boldsymbol v_c\boldsymbol\cdot\mathsf I^s+\mathsf I^s\boldsymbol\cdot(\boldsymbol\nabla^s\otimes \boldsymbol v_c)^T] - \alpha c \mathsf I^s.
\end{equation}
Here, $\alpha$ is the 2D bulk modulus of the cortex.
This results in an additional term in the cortex force balance which accounts for an elastic force.
The linear stability analysis follows the procedure described above in section \ref{sec_linear_stability_supplement}.
From this we obtain an evolution equation for the amplitude of the concentration as in Eq.\ \eqref{eq_c_evolution_SI} with the modified growth rates
\begin{equation}
	\label{lambdal_compress}
	\lambda_l = \frac{2l(l+1)(\eta_s+\eta_b)}{(2\eta_s+\eta_b)l(l+1)-2\eta_s}\frac{v^0_p c_0}{R_0 c_r} e^{-\frac{c_0}{c_r}}- l(l+1) \frac{D}{R_0^2}  - \beta - \frac{\frac{l}{2} (l+1) }{ (l^2 + l -1) \eta_s + \frac{l}{2}(l+1) \eta_b } \alpha c_0.
\end{equation}
Compared to the expression in Eq.\ \eqref{lambdal}, the here obtained growth rates contain a correction which is proportional to the stiffness $\alpha$ and decreases $\lambda_l$. 
Thus, the passive stress caused by the elastic resistance of the actin meshwork counter-acts the described instability.
The onset of cell polarity takes place for $\lambda_1 > 0$ which is the case for $v_p^0 > v_p^c$, with the critical polymerization speed
\begin{equation}
	\label{eq_vpc_compress}
	v_p^c=\frac{c_r}{c_0} e^\frac{c_0}{c_r} \left(\frac{D}{R_0}+\frac{\beta R_0}{2} + \frac{\alpha c_0 R_0}{ \eta_s + \eta_b}\right) \,.
\end{equation}
The elastic shear modulus of the F-actin network, $G$, is at the order of several hundreds of Pascals \cite{janmey1994mechanical, doi:10.1126/science.1095087}. Assuming $G = \SI{300}{\pascal}$ \cite{doi:10.1126/science.1095087} and $h = \SI{190}{\nano \meter}$ \cite{clark2013monitoring}, we estimate $\alpha = G h / c_0 = \SI{1.4e-19}{\pascal \meter \cubed}$.
With this we obtain from Eq.\ \eqref{eq_vpc_compress} and the parameters used in section \ref{sec_parameters} a critical polymerization speed of $v_p^c = \SI{1.77}{\micro \meter \per \second}$. 
This represents an $21 \%$ increase of the instability threshold compared to the $\alpha = 0$ case (see section \ref{sec_parameters}) and lies within the range of typical polymerization speeds reported in the literature.

\section{Weakly nonlinear analysis including inverse relation for the polymerization velocity}
\label{sec_weakly_nonlinear_inverse}
In the following we perform a weakly nonlinear analysis assuming an inverse relation between the polymerization velocity $\bm v_p$ and the actin concentration $c$,
\begin{equation}\label{eq_v_c_inverse}
	\bm v_p = v_p^0 \frac{c_r}{c} \bm n
\end{equation}
where $c_r$ is a constant. This is in alignment with the assumption that filament growth is slowed by a competition for actin monomers (see section \ref{sec_estimate_monomer_competition}).
The calculation follows the procedures described above in section \ref{sec_nonlinear_analysis_supplement}.
With the actin concentration given by Eq.\ \eqref{eq_init_conc_second_order}, we obtain from Eq.\ \eqref{eq_v_c_inverse} the polymerization velocity up to third order in $\varepsilon$,
\begin{equation}\label{eq_actin_vp_higher_order_inverse}
	\bm v_p (\theta, \varphi) = v_p^0 \frac{c_r}{c_0} \Biggl\{ 1 - \varepsilon \frac{\delta \hat c_1}{c_0} \cos \theta + \frac{\varepsilon^2}{2}  \left[ \frac{\delta \hat c_2}{c_0} + \cos^2 \theta \left( \frac{\delta \hat c_1^2}{c_0^2} - 3 \frac{\delta \hat c_2}{c_0} \right) \right] - \frac{\varepsilon^3}{6} \frac{\delta \hat c_1}{c_0} \cos \theta \left[ \left(  \frac{\delta \hat c_1^2}{c_0^2} - 9 \frac{\delta \hat c_2}{c_0} \right) \cos^2 \theta + 3 \frac{\delta \hat c_2}{c_0} \right] \Biggr\} \bm e_r,
\end{equation}
By solving the force balance equation under the condition of a fixed cell shape for $l = 2$, we obtain the shape coefficient
\begin{equation}\label{eq_rho_2_inverse}
	\varrho_2 = 4 \varepsilon^2 v_p^0 c_r \frac{  \eta_s (\eta_s + \eta_b) \left( 2 \frac{\delta \hat c_1^2}{c_0^2} - 3 \frac{\delta \hat c_2}{c_0} \right) }{3 c_0 \Delta P R_0^2 \left( 5 \eta_s + 3\eta_b \right)}.
\end{equation}
With this we obtain from the advection-diffusion equation the time-evolution of the amplitudes of the concentration harmonics,
\begin{align}
	\label{eq_coeff_cdot_actin_non_linear_0_inv}
	\dot{c}_0 &= 0, \\
	\label{eq_coeff_cdot_actin_non_linear_1_inv}
	\delta \dot{\hat c}_1 &= \varepsilon \lambda_1 \delta \hat c_1 + \varepsilon^3  \alpha_1 \delta \hat c_1^3 + \varepsilon^3 \beta_1  \delta \hat c_1 \delta \hat c_2 , \\
	\label{eq_coeff_cdot_actin_non_linear_2_inv}
	\delta \dot{\hat c}_2 &= \varepsilon^2  \lambda_2 \delta \hat c_2 + \varepsilon^2 \alpha_2 \delta \hat c_1^2 .
\end{align}
Herein, the linear growth rates are given by
\begin{equation}
	\label{eq_linear_growth_rate_inv}
	\lambda_1 = 2 \frac{v^0_p c_r}{R_0 c_0}- 2 \frac{D}{R_0^2}  - \beta, \ \qquad
	\lambda_2 = 6 \frac{v^0_p c_r (\eta_s + \eta_b)}{R_0 c_0 (5 \eta_s + 3 \eta_b)}  - 6 \frac{D}{R_0^2} - \beta
\end{equation}
and the nonlinear coefficients by
\begin{equation}
	\label{eq_nonlinear_inv}
	\alpha_1 = \frac{2 v^0_p c_r (13 \eta_s + 7 \eta_b)}{5 R_0 c_0^3 (5 \eta_s + 3 \eta_b)},\ \qquad
	\beta_1 = - \frac{4 v^0_p c_r (11 \eta_s + 6 \eta_b)}{5 R_0 c_0^2 (5 \eta_s + 3 \eta_b)}  , \qquad	\alpha_2 = \frac{2 v^0_p c_r (3 \eta_s + \eta_b)}{R_0 c_0^2 (5 \eta_s + 3 \eta_b)} .
\end{equation}
Note that the linear growth rates in Eq.\ \eqref{eq_linear_growth_rate_inv} correspond up to prefactors in the $v_p^0$ terms to the expressions obtained for the exponential velocity-concentration relation in Eqs.\ \eqref{lambda1} and \eqref{lambdal} (see also linear terms in Eqs.\ \eqref{eq_coeff_cdot_actin_non_linear_1} and \eqref{eq_coeff_cdot_actin_non_linear_2}).
By setting $\lambda_1 = 0$ in Eq.\ \eqref{eq_linear_growth_rate_inv} we obtain the critical polymerization speed
\begin{equation}
	\label{vpc_inv}
	v_p^c=\frac{c_0}{c_r}\left(\frac{D}{R_0}+\frac{\beta R_0}{2}\right)\,.
\end{equation}
The condition $\delta \dot{\hat c}_2 = 0$ and Eq.\ \eqref{eq_coeff_cdot_actin_non_linear_2_inv} allows us to eliminate $\delta \hat c_2$ in Eq.\ \eqref{eq_coeff_cdot_actin_non_linear_1_inv}. Upon evaluating the nonlinear coefficients in Eq.\ \eqref{eq_nonlinear_inv} at $v_p^0 = v_p^c$, we obtain an evolution equation for the amplitude of the first concentration harmonic,
\begin{equation}\label{eq_first_harmonic_third_order_inv}
	\delta \dot{\hat c}_{1} = \frac{2 c_r }{R_0 c_0} (v_p^0 - v_p^c) \delta \hat c_1 + \nu \delta \hat c_1^3,
\end{equation}
with the cubic coefficient
\begin{equation}\label{eq_non_linear_coefficient_inv}
\nu = \frac{2 (2D + \beta R_0^2) \left[ D(9\eta_s + 5 \eta_b) - \beta R_0^2 (2 \eta_s + \eta_b) \right]}{5 R_0^2 c_0^2 \left[ 6 D (2 \eta_s + \eta_b) + R_0^2 \beta \eta_s \right]}.
\end{equation}
From Eq.\ \eqref{eq_actin_vp_higher_order_inverse} we obtain the $\mathcal{O} (\varepsilon)$ contribution to the cortex velocity which describes the transformation to the cell frame,
\begin{equation}\label{eq_vc_inv}
	\boldsymbol{v}_c = \varepsilon  \frac{v_p^0 c_r \delta \hat{c}_1}{c_0^2} \boldsymbol{e}_z.
\end{equation}
The final expression for the retrograde flow velocity at the equator of the cell, i.e., the negative cell migration velocity, is then obtained by substituting the non-trivial steady-sate solution of Eq.\ \eqref{eq_first_harmonic_third_order_inv} for $\delta \hat{c}_1$ into Eq.\ \eqref{eq_vc_inv},
\begin{equation}\label{eq_vMig_inv}
	\bm v_0  = \frac{ v_p^c c_r}{c_0^2}  \sqrt{\frac{2 c_r (v_p^0 - v_p^c)}{- R_0 c_0 \nu}} \boldsymbol{e}_z.
\end{equation}
The retrograde flow speed $v_0 = |\bm v_0|$ according to Eq.\ \eqref{eq_vMig_inv} is shown in Fig.\ \ref{fig_compare_inverse_law} together with the result obtained above in section \ref{sec_nonlinear_analysis_supplement} for the exponential polymerization law (see also analytical branch in Fig.\ 3A of the main text).
In both cases $\nu < 0$ is obtained which implies a continuous onset of motility (supercritical bifurcation). Note that the critical polymerization speed obtained for the inverse polymerization law is reduced compared to $v_p^c$ obtained with the exponential relation.
\begin{figure}[h]\centering
	\includegraphics[width=0.75\textwidth]{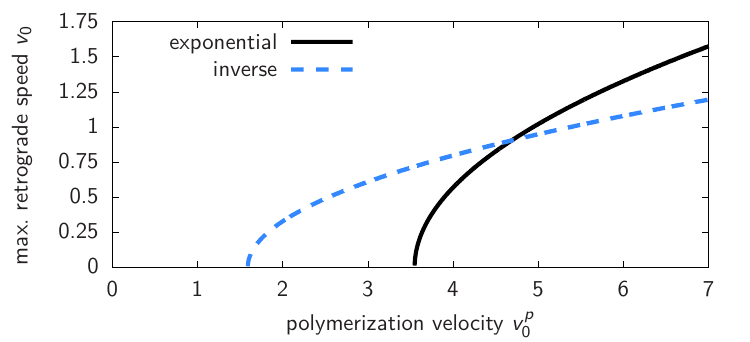}
	\caption{{Maximal retrograde flow speed (cell migration speed) as a function of the polymerization speed. }}
	\label{fig_compare_inverse_law}
\end{figure}
}